\documentclass{article}
\usepackage{arxiv}
\usepackage{graphicx,amsmath,amsfonts,amscd,amssymb,pstricks}
\usepackage{color}% Include figure files
\usepackage{dcolumn}% Align table columns on decimal point
\usepackage{bm}% bold math
\usepackage{longtable}
\usepackage{mathrsfs}
\usepackage{textcomp}
\usepackage{esvect}
\usepackage{float}
\usepackage{mathtools,amsfonts}
\usepackage[USenglish,british]{babel}
\usepackage{environ}
\usepackage{xifthen}
\usepackage{xargs}		% programming: better newcommand
\usepackage{hyperref}
\usepackage{physics,mathtools}
\usepackage[separate-uncertainty = true]{siunitx}
\usepackage{multirow}
\usepackage{comment}
\usepackage{enumerate}
\usepackage{appendix}
\usepackage{tikz}	
\usetikzlibrary{arrows, calc, matrix, patterns, decorations.markings, shapes, decorations.pathmorphing, shadows.blur}
\usepackage[utf8]{inputenc}
\usepackage{enumitem}
\usepackage{xcolor}

\title{Probing the diffusive behaviour of beam-halo dynamics in circular accelerators}
\author{C.\ E.\ Montanari\\
Physics and Astronomy Department, Bologna University and INFN, Bologna, Italy \\
Beams Department, CERN, Esplanade des Particules 1, 1211 Meyrin, Switzerland\\
\And
A.\ Bazzani\\
Physics and Astronomy Department, Bologna University and INFN, Bologna, Italy \\
\And
M.\ Giovannozzi\thanks{Corresponding author: massimo.giovannozzi@cern.ch}
Beams Department, CERN, Esplanade des Particules 1, 1211 Meyrin, Switzerland}% etc
\begin{document}
\maketitle
\begin{abstract}
Circular particle accelerators at the energy frontier are based on superconducting magnets that are extremely sensitive to beam losses as these might induce quenches, i.e.\ transitions to the  normal-conducting state. Furthermore, the energy stored in the circulating beam is so large that hardware integrity is put in serious danger, and machine protection becomes essential for reaching the nominal accelerator performance. In this challenging context, the beam halo becomes a potential source of performance limitations and its dynamics needs to be understood in detail to assess whether it could be an issue for the accelerator. In this paper, we discuss in detail a novel framework, based on a diffusive approach, to model beam-halo dynamics. The functional form of the optimal estimate of the perturbative series, as given by Nekhoroshev's theorem, is used to provide the functional form of the action diffusion coefficient. The goal is to propose an effective model for the beam-halo dynamics and to devise an efficient experimental procedure to obtain an accurate measurement of the diffusion coefficient.
\end{abstract}
%
%\PACS{
%      {PACS-key}{discribing text of that key}   \and
%      {PACS-key}{discribing text of that key}
%     } % end of PACS codes
%} %end of abstract
%
%%%%%%%%%%%%%%%%%%%%%%%%%%%%%%%%%%%%%%%%%%%%%%%%%%%%%%%%%%%%%%%%%%%%%%%%%%%%%%%%

\section{Introduction}
\label{sec:introduction}

%%%%%%%%%%%%%%%%%%%%%%%%%%%%%%%%%%%%%%%%%%%%%%%%%%%%%%%%%%%%%%%%%%%%%%%%%%%%%%%%

For the design and operation of modern circular particle accelerators, understanding the complex dynamics that characterises the beam-halo formation and evolution is of paramount importance. Indeed, several phenomena leading to particle loss and beam-quality degradation, crucial to determine the performance of a particle accelerator, are closely linked to the evolution of the beam halo. 

This is particularly true for present and future colliders based on superconducting magnets, such as LHC~\cite{LHCDR}, its upgrade HL--LHC~\cite{BejarAlonso:2749422}, or the proposed FCC-hh~\cite{FCC-hhCDR}. Beam losses have a direct impact on the accelerator performance. 

Beam-halo dynamics is governed by a multitude of effects, such as the unavoidable non-linear field errors of the superconducting magnets, as well as ripples in the magnets' power converters. In general, the beam dynamics of hadron machines is accurately described in terms of a Hamiltonian from which the equations of motion can be derived. If the system under consideration includes time-dependent effects, this turns into a radical change of the character of the beam dynamics. For instance, the presence of modulation of the characteristic frequencies of the Hamiltonian system implies the existence of extended weakly-chaotic layers in the phase space~\cite{NEISHTADT1991}. In these regions, it is possible to model the orbit diffusion by a stochastic process. The situation worsens in case the periodic modulations themselves resemble stochastic processes since the diffusive behaviour might involve the whole of the accessible phase space.

Recently, a framework has been developed and proposed~\cite{Bazzani:2019lse,bazzani2020diffusion}, in which the long-term behaviour of the beam dynamics and particle losses in circular accelerators is described by means of a diffusive model. In this framework, the evolution of the beam distribution can be described by a Fokker-Planck (FP) equation, in which the diffusion coefficient represents the key quantity to describe the beam dynamics. The development of diffusive models of the transverse dynamics of charged particles is not at all new for accelerator physics, and a rather broad literature exists (see , e.g.\ Refs~\cite{PhysRevLett.68.33,gerasimov1992applicability,MESS1994279,zimmermann1994transverse,PhysRevLett.77.1051,PhysRevSTAB.5.074001,stancari2011diffusion,PhysRevSTAB.15.101001} and references therein). However, the model that we developed has a very peculiar feature, since we assume that the functional form of the diffusion coefficient is derived from the optimal estimate of the perturbation series provided by the Nekhoroshev's theorem~\cite{Nekhoroshev:1977aa,Bazzani:1990aa,Turchetti:1990aa}. 

The FP equation is suitable to study the evolution of a beam distribution in the presence of collimators, whose jaws can be represented by the absorbing boundary conditions needed to solve the FP equation. Furthermore, so-called collimators' scans can be used to probe the beam-halo dynamics and, in particular, to reconstruct the behaviour of the diffusion coefficient as a function of transverse amplitude~\cite{MESS1994279,stancari2011diffusion,PhysRevSTAB.16.021003,PhysRevAccelBeams.23.044802}. The method of collimator scans has been intensively used at the LHC: it is based on small displacements of the jaws combined with the measurement of the beam losses. The displacements can be either inward or outward, and depending on the direction, the local losses feature different behaviour. The interpretation of the experimental data relies on a number of assumptions that are closely linked to the form of the FP equation that is used to model the beam dynamics. 

The special functional form of the diffusion coefficient that we have proposed requires the study of an appropriate protocol to probe its properties by means of a beam test and hopefully, it shall bring a useful additional method to describe the beam-halo dynamics. In this paper, the properties of the FP equation, in particular that of the outgoing current at a boundary condition, are studied in detail by means of analytical models and even more by means of numerical simulations. These analyses lead to the definition of an optimal protocol to extract the information about the diffusion coefficient by performing a sequence of well-chosen variations of the position of the boundary condition. An important part of our approach concentrates on the determination of the accuracy and robustness of the proposed protocol.

The plan of the paper is the following. In Section~\ref{sec:the_diffusion_approach}, we present the theoretical framework that defines the Nekhoroshev-like diffusive model, along with some considerations on the special form of the diffusion coefficient and its implications on the dynamics. In Section~\ref{sec:some_considerations}, we analyse the main characteristics of a FP process, with focus on the outgoing current, its behaviour in various conditions, such as in stationary or semi-stationary equilibrium. In Section~\ref{sec:moving_the_absorbing_barrier}, we discuss how the outgoing currents obtained from outward or inward changes of the position of the boundary condition can be described by our model, and how such currents can be split in two processes with distinct timescales. In Section~\ref{sec:numerical_results}, the results presented in the previous sections are used to define a protocol to reconstruct the diffusion coefficient of the FP process. The results of detailed numerical simulations are presented and discussed, quantifying the performance of the proposed method. Finally, in Section~\ref{sec:conclusions} some conclusions are drawn, whereas detail about the numerical integration of the FP process is discussed in Appendix~\ref{app_sec:numerical_integration_with_crank_nicolson} and some analytical computations are presented in the Appendixes~\ref{app_sec:analytic_estimate_of_the_current_loss} and~\ref{app_sec:outgoing_current_for_a_system_with_infinite_source}.

%%%%%%%%%%%%%%%%%%%%%%%%%%%%%%%%%%%%%%%%%%%%%%%%%%%%%%%%%%%%%%%%%%%%%%%%%%%%%%%%

\section{The diffusion approach to non-linear dynamics}
\label{sec:the_diffusion_approach}

%%%%%%%%%%%%%%%%%%%%%%%%%%%%%%%%%%%%%%%%%%%%%%%%%%%%%%%%%%%%%%%%%%%%%%%%%%%%%%%%

From Hamiltonian systems theory, we know that orbit diffusion in the phase space is related to the presence of extended weakly-chaotic regions~\cite{froeschle1999weak}. Otherwise, the presence of invariant Kolmogorov--Arnol'd--Moser tori ensures long-term stability of the dynamics~\cite{morbidelli1995connection}.

In circular particle accelerators, the transverse dynamics is affected by a multitude of unavoidable small random perturbations~\cite{ellison1999noise}, as well as slow modulation of magnet currents and transverse tune ripples that could lead to the formation of these weakly-chaotic regions. Therefore, it is reasonable to assume that the particle motion is described by models of the form 
\begin{equation}
H(\theta, I, t)=H_{0}(I)+ \xi(t) H_{1}(\theta, I) \, , 
\label{hstoc}
\end{equation}
where $(I, \theta)$ are action-angle variables and $\xi(t)$ is a continuous stationary stochastic noise with zero mean that represents the effect of the chaotic dynamics. In the case of a small stochastic perturbation if the correlation time scale of the noise $\xi(t)$ is much shorter than the evolution time scale of the unperturbed system, it is possible to describe the evolution of the action density function $\rho(I,t)$ by the FP equation (see, e.g.\ Ref.~\cite{bazzani2020diffusion} for the mathematical details and Ref.~\cite{Montanari:2728138} for an application to a stochastic symplectic map)
\begin{equation}
    \frac{\partial \rho (I, t)}{\partial t}=\frac{1}{2} \frac{\partial}{\partial I} {D}(I) \frac{\partial}{\partial I}\, \rho(I, t) \, ,
    %\frac{\partial \rho (I, t)}{\partial t}=\frac{\varepsilon^{2}}{2} \frac{\partial}{\partial I} {D}(I) \frac{\partial}{\partial I}\, \rho(I, t) \, , 
    \label{eq:fp}
\end{equation}
where the diffusion coefficient reads
\begin{equation}
D(I)=\sqrt{\left \langle \frac{\partial H_1}{\partial \theta} \right \rangle^2}
\label{qsapp}
\end{equation}    
and the operator $\langle\,\rangle$ denotes the average value with respect to the angle variables. The stochastic perturbation in Eq.~\eqref{hstoc} models the effect
of a large weakly-chaotic region in the phase space and its amplitude should be related to the non-integrability of the dynamics. We assume, therefore, that its amplitude is related to the functional form provided by the optimal estimate of the perturbative series, according to the Nekhoroshev's theorem~\cite{Turchetti:1990aa,Bazzani:1990aa}. Therefore, the diffusion coefficient has the form
\begin{equation}
    %D(I)=c \exp \left[-2\left(\frac{I_\ast}{I}\right)^{\frac{1}{2\kappa}}\right]\, , \qquad \int_{0}^{I_{\mathrm{a}}} D(I) \, \mathrm{d}I = 1 \, , 
    D(I)\propto\exp \left[-2\left(\frac{I_\ast}{I}\right)^{\frac{1}{2\kappa}}\right]\, , 
    \label{eq:diffusion}
\end{equation}
and the two parameters $\kappa, I_\ast$ in Eq.~\eqref{eq:diffusion} have a physical meaning that stems from the Nekhoroshev's theorem: the exponent $\kappa$ is related to the analytic structure of the perturbative series and the dimensionality of the system, and is supposed to be independent from the intensity of the perturbation themselves~\cite{bazzani2020diffusion}; $I_\ast$ is related to the asymptotic character of the perturbative series.

In the following, we consider 1DoF Hamiltonian systems and we introduce an absorbing boundary condition at $I_{\mathrm{a}}$, i.e.\ the phase-space limit beyond which an initial condition is considered lost. Note that $D(I)$ and $\rho$ have dimensions $[I^2t^{-1}]$ and $[I^{-1}]$, respectively.

\begin{figure*}[htp]
    \centering
    \includegraphics{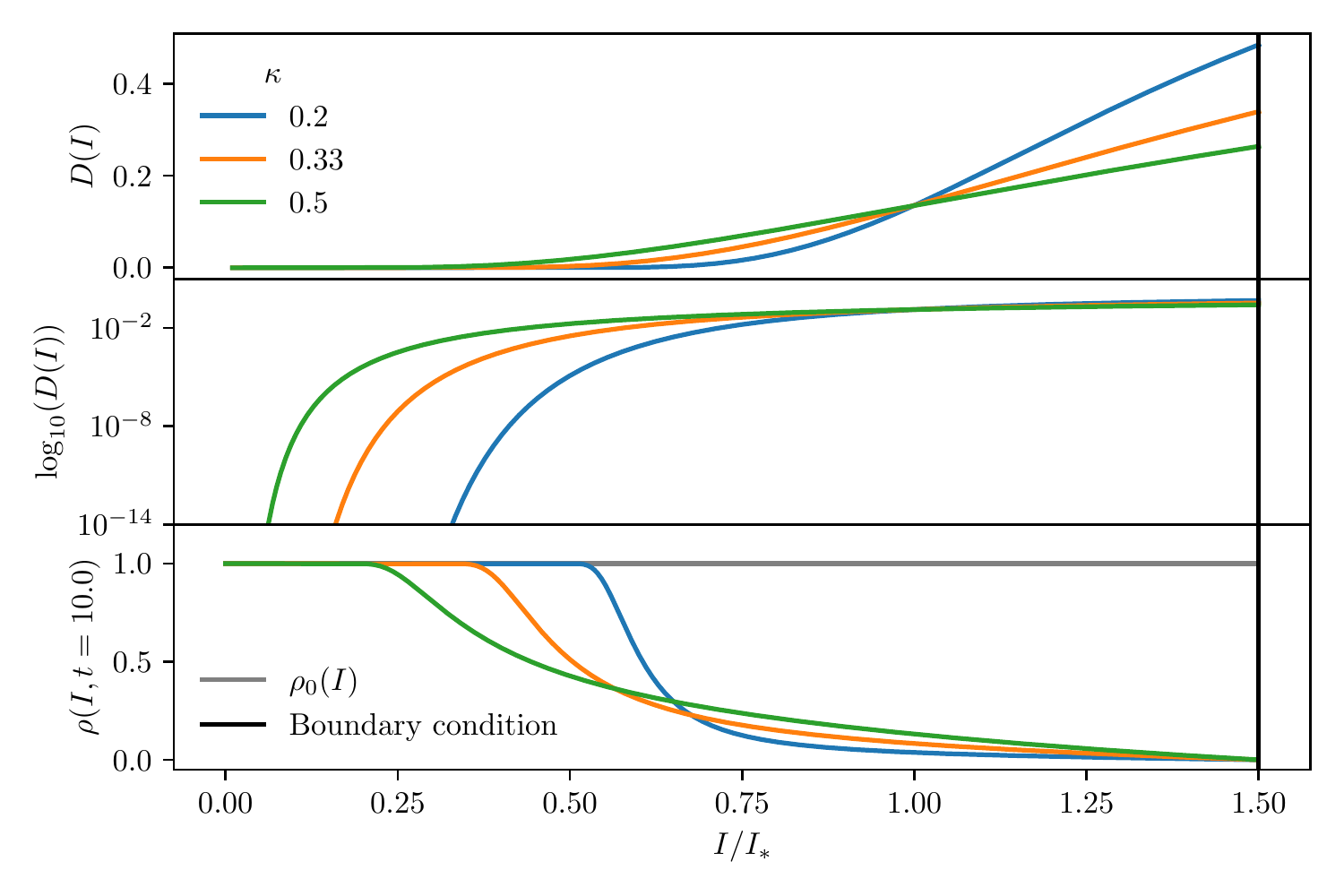}
    \caption{Top and centre: plot of $D(I)$ both in linear and logarithmic scale for three values of $\kappa$ as a function of $I/I_\ast$.
    %For the cases shown here, the choice $c=1$ has been made, to have comparable curves.
    Bottom: evolution of an uniform distribution (in grey) over the same time interval corresponding to $(t=10.0 \, [\text{a. u.}])$ for three values of $\kappa$. (Simulations parameter: $(I_\ast = 1.0\,[\sigma^2])$).}
    \label{fig:1}
\end{figure*}

In Fig.~\ref{fig:1} (top and centre), we consider the behaviour of $D(I)$ from Eq.~\eqref{eq:diffusion} for some values of $\kappa$. We can distinguish three regions for this type of function: $(i)$ a stable-core region for $I \ll I_\ast$, for which $D(I)$ has values decreasing to zero exponentially fast; $(ii)$ a ramp-up region for $I \lesssim I_\ast$, where $D(I)$ starts to have non-negligible values and changes from an exponential growth to an almost linear one; $(iii)$ a region for $I >  I_\ast$, where $D(I)$ features an almost linear growth (in logarithmic scale a saturation appears).

These three regions are more or less distinguishable depending on the value of $\kappa$. In Ref.~\cite{bazzani2020diffusion}, a value of $\kappa$ around $0.33$ was found as the best fit to the data measured during experimental studies at the LHC, and for this reason this value is used in the numerical simulations presented in this study. 

In Fig.~\ref{fig:1} (bottom) we also display the result of the numerical integration of Eq.~\eqref{eq:fp} for some values of $\kappa$, performed on an initial distribution $\rho_0(I)=1$ with a Crank-Nicolson scheme~\cite{crank1947practical} (see Appendix~\ref{app_sec:numerical_integration_with_crank_nicolson} for some detail on the integration scheme used in our studies).
%, using $\varepsilon^2 = 1$ in all numerical simulations performed, and setting the absorbing boundary condition at $I/I_\ast=1.5$. 
It is also possible to observe in the shape of the distribution function a stable-core region, corresponding to $I/I_\ast \ll 1$, where $D(I)$ starts having values very close to zero, a fast decrease region and, finally, a saturation region, for $I/I_\ast = 1$ and beyond, where $\rho(I, t)$ assumes small values.

We treat our problem by using the 1D action variable $I$, representing the non-linear invariant of the system. We consider the rescaled action variable $I \to I/\sigma^2$, and express the action in units of beam emittance, and this action will therefore be a dimensionless quantity.

As for the initial condition for the beam distribution, we use the exponential distribution
\begin{equation}
    \rho_0(I) = \exp(-I) \, , 
    \label{eq:initial_distribution}
\end{equation}
obtained by the transformation of the standard Gaussian distribution in physical variables. We also note that for future analysis, it might be interesting to consider beam distributions made of combinations of exponential distributions, as this could be used to simulate the behaviour of a beam with overpopulated tails.

%%%%%%%%%%%%%%%%%%%%%%%%%%%%%%%%%%%%%%%%%%%%%%%%%%%%%%%%%%%%%%%%%%%%%%%%%%%%%%%%
\section{Some considerations on FP processes}\label{sec:some_considerations}

\subsection{Outgoing current}\label{subsec:outgoing_current}

%%%%%%%%%%%%%%%%%%%%%%%%%%%%%%%%%%%%%%%%%%%%%%%%%%%%%%%%%%%%%%%%%%%%%%%%%%%%%%%%

In a generic diffusive process, the outgoing current at the absorbing boundary condition at $I_\mathrm{a}$ is defined as
\begin{equation}
    J_\mathrm{a}(t) = D(I_\mathrm{a})\pdv{\rho(I, t)}{I} \Bigr|_{(I_\mathrm{a}, t)}\,.
    \label{eq:outgoing_current_definition}
\end{equation}
Equation~\eqref{eq:fp} provides a means to obtain an analytical estimate of the current lost at the absorbing barrier (see Appendix~\ref{app_sec:analytic_estimate_of_the_current_loss} for the mathematical details). We first consider the change of variable
\begin{equation}
    x(I) = -\int_I^{I_\mathrm{a}} \frac{1}{D^{1/2}(I')}\,\mathrm{d}I' \,, \quad \rho_x(x,t) = \rho(I, t) \sqrt{D(I)} \,, \quad x_\mathrm{a}=x(I_\mathrm{a})=0 \, ,
    \label{eq:change_of_variable}
\end{equation}
we can consider the FP problem in the Smoluchowsky form
\begin{equation}
    \pdv{\rho_x}{t} = \frac{1}{2}\pdv{x} \dv{V(x)}{x} \rho_x+\frac{1}{2}\pdv[2]{\rho_x}{x} \quad \text{where} \quad V(x) = -\ln\left(D^{1/2}(x)\right) \, .
    \label{eq:smol_and_potential}
\end{equation}

If we then consider the linearisation of the potential $V(x)$ at $x_0$ in the form $-\nu \, x$, the following expression for the outgoing current at  $x_a = 0$ for an initial distribution $\delta(x - x_0)$ is obtained
\begin{equation}
    J_\mathrm{a}(x_0, t) = \frac{|x_0|}{t\sqrt{2\pi t}}\exp\left(-\frac{(x_0+\frac{\nu}{2}t)^2}{2t}\right) \,,
    \label{eq:thecurrent}
\end{equation}
where $\nu$, the linearisation of the potential~\eqref{eq:smol_and_potential} at $x_0$ with $D(I)$ given by Eq.~\eqref{eq:diffusion} in the new coordinates, has the following expression
\begin{equation}
    \nu=\frac{1}{2\kappa}\frac{1}{I(x_0)}\left(\frac{I_\ast}{I(x_0)}\right)^{\frac{1}{2\kappa}}\exp\left[-\left(\frac{I_\ast}{I(x_0)}\right)^{\frac{1}{2\kappa}}\right]\,.
    \label{eq:linearised}
\end{equation}

We remark that Eq.~\eqref{eq:thecurrent} can be applied to a generic distribution $\rho_0$ via a convolution
\begin{equation}
    J_\mathrm{a}(t) = \int J_\mathrm{a}(x,t)\rho_x(x)\,\mathrm{d}x\,.
    \label{eq:current_convolution}
\end{equation}
We also note that Eqs~\eqref{eq:thecurrent} and~\eqref{eq:linearised} provide inevitably an underestimate of the actual current lost~\cite{montanari:ipac2021:tupab233}, as the actual drift term is a positive increasing function for $I\ll I_\ast$. However, we expect a good description of the local behaviour close to the absorbing boundary condition, i.e.\ we obtain a good estimate of the current lost for initial distributions that are close enough to the absorbing barrier at $I=I_\mathrm{a}$. 

%%%%%%%%%%%%%%%%%%%%%%%%%%%%%%%%%%%%%%%%%%%%%%%%%%%%%%%%%%%%%%%%%%%%%%%%%%%%%%%%

\subsection{Stationary system with a constant  source}\label{subsec:system_infinite_regime}

%%%%%%%%%%%%%%%%%%%%%%%%%%%%%%%%%%%%%%%%%%%%%%%%%%%%%%%%%%%%%%%%%%%%%%%%%%%%%%%%

Let us consider a diffusive process within the region $[I_0, I_\mathrm{a}]$, with an absorbing boundary condition $\rho(I_\mathrm{a}, t) = 0$, and $\rho(I_0, t) = 1$ as a constant source over time. Regardless of the shape of the initial distribution $\rho_0$, the system will eventually relax to its equilibrium distribution $\rho_\text{eq}(I)$, characterised by a constant outgoing current at $I_\mathrm{a}$. Such a distribution satisfies the following equation
\begin{equation}
    \pdv{I} D(I) \pdv{I} \rho_{\text{eq}}(I) = 0 \,,
\end{equation}
whose solution is given by
\begin{equation}
    \rho_\text{eq}(I) = \alpha \int_I^{I_\mathrm{a}} \frac{1}{D(x)}\,\mathrm{d}x \,, \quad \alpha = \frac{1}{ \displaystyle{\int_{I_0}^{I_\mathrm{a}} \frac{1}{D(x)}\,\mathrm{d}x}} \, .
    \label{eq:equilibrium_stationary_distribution}
\end{equation}

The relaxed system features a constant outgoing current given by
\begin{equation}
    J_\mathrm{a}(t) = D(I_\mathrm{a})\pdv{\rho_\text{eq}}{I}\Bigr|_{(I_\mathrm{a},t)} = \alpha\,,
    \label{eq:equilibrium_stationary_current}
\end{equation} 
which is directly linked to the integral of the diffusion coefficient. For a Nekhoroshev-like diffusion coefficient, we have the analytical expression
\begin{equation}
    \rho_\text{eq}(I) = \alpha \int_I^{I_\mathrm{a}} \exp\left[2\left(\frac{I_\ast}{x}\right)^{\frac{1}{2\kappa}}\right] \,\mathrm{d}x 
    = 2\alpha\ \kappa x \left[-2\left(\frac{I_\ast}{x}\right)^{\frac{1}{2\kappa}}\right]^{2\kappa} \Gamma\left(-2\kappa, -2\left(\frac{I_\ast}{x}\right)^{\frac{1}{2\kappa}}\right)\Bigg|_I^{I_\mathrm{a}}\,,
\end{equation}
where $\Gamma$ is the upper incomplete gamma function
\begin{equation}
    \Gamma(s,x) = \int_x^{\infty} t^{s-1}\,e^{-t}\,{\rm d}t \,.
\end{equation}

When the system is out of equilibrium, one can obtain an analytical description of the outgoing current by using the formula in Eq.~\eqref{eq:thecurrent}, where, instead of performing a convolution between Eq.~\eqref{eq:thecurrent} and $\rho_0$, we perform a convolution with $\rho_0 - \rho_\text{eq}$. The resulting outgoing current is added to the constant value $\alpha$ (the mathematical details of such procedures are illustrated in Appendix~\ref{app_sec:outgoing_current_for_a_system_with_infinite_source}).

%%%%%%%%%%%%%%%%%%%%%%%%%%%%%%%%%%%%%%%%%%%%%%%%%%%%%%%%%%%%%%%%%%%%%%%%%%%%%%%%

\subsection{Semi-stationary regime for a real system}
\label{subsec:semi_stationary_regime_for_a_real_system}

%%%%%%%%%%%%%%%%%%%%%%%%%%%%%%%%%%%%%%%%%%%%%%%%%%%%%%%%%%%%%%%%%%%%%%%%%%%%%%%%

When working with a Nekhoroshev-like diffusion coefficient, its exponentially small values for $I \ll I_\ast$ generate a stable-core region with extremely low diffusion rates (see Fig.~\ref{fig:1}). This observation can be shown by computing the time of the maximum of the outgoing current for an initial distribution $\delta(I - I_0)$. We consider the time derivative of Eq.~\eqref{eq:thecurrent}
\begin{equation}
\pdv{J_\mathrm{a}(x_0, t)}{t} = \frac{\sqrt{2} x_0 \left[12 t + \left(\nu t - 2 x_0\right) \left(\nu t + 2 x_0\right)\right] }{16 \sqrt{\pi t^7}}\exp\left(-\frac{(x_0+\frac{\nu}{2}t)^2}{2t}\right)\,,
\end{equation}
which is zero for two values of $t$ of opposite sign, and the positive one is
\begin{equation}
    t_{\text{max}}(x_0) = \frac{2 \left(\sqrt{\nu^{2} x_0^{2} + 9} - 3\right)}{\nu^{2}}\,.
    \label{eq:taumax}
\end{equation}
Considering the diffusion coefficient of Eq.~\eqref{eq:diffusion} in the change of variable Eq.~\eqref{eq:change_of_variable}, we have that
\begin{equation}
    x_0(I_0) = \int_{I_0}^{I_\mathrm{a}} \exp\left[\left(\frac{I_\ast}{I}\right)^{\frac{1}{2\kappa}}\right]\,\mathrm{d}I\,.
    \label{eq:peak_current_time}
\end{equation}

We observe that the modulus of the integral of Eq.~\eqref{eq:peak_current_time} increases exponentially for $I_0 \ll I_\ast$. Likewise, the value of $\nu$, given in Eq.~\eqref{eq:linearised}, decreases exponentially in the same range of values of $I_0$, which characterise a strong exponential variation for $t_\text{max}$ as a function of $I_0$. This fact suggests that the contribution to the outgoing current at an absorbing barrier at time $t$ is mainly determined by the initial conditions near $I_0$, with $t_\text{max}(I_0) \approx t$. Hence, given a generic initial distribution $\rho_0(I)$ and an absorbing boundary condition at $I_\mathrm{a}$, after a transient time $t$, the system relaxes to a condition where the current $J_\mathrm{a}(t)$ is mainly determined by $\rho_0(I_0)$, where $I_0$ satisfies $t_\text{max}(I_0) \approx t$. Considering the exponential increase of $t_\text{max}(I)$, we have a core region that is slowly eroded by the diffusive process. Outside of this core region, the system behaves as in a semi-stationary regime, characterised by a very slowly varying source at $I_0$.

The evolution of this semi-stationary process can be approximated by modifying the $\alpha$ term in Eq.~\eqref{eq:equilibrium_stationary_distribution} as
\begin{equation}
    \rho_\mathrm{eq}(I, t) = \alpha(t) \int_I^{I_\mathrm{a}} \frac{1}{D(x)}\,\mathrm{d}x\,,
    \label{eq:semi_stationary_distribution}
\end{equation}
where here $\alpha(t)$ depends on the value of the initial distribution $\rho_0(I_0)$, and can be estimated by
\begin{equation}
    \alpha(t) = \frac{\rho_0\left(I_0(t)\right)}{\displaystyle{ \int_{I_0(t)}^{I_\mathrm{a}} \frac{1}{D(x)}\,\mathrm{d}x}}\,,
    \label{eq:alpha_with_time}
\end{equation}
in which $I_0(t)$ is obtained by inverting Eq.~\eqref{eq:peak_current_time} to determine $I_0(x_0)$, and Eq.~\eqref{eq:taumax} to obtain $x_0(t_\mathrm{max})$. By composing the two functions, $I_0(t)$ is determined.

\begin{figure*}[htp]
    \centering
    \includegraphics{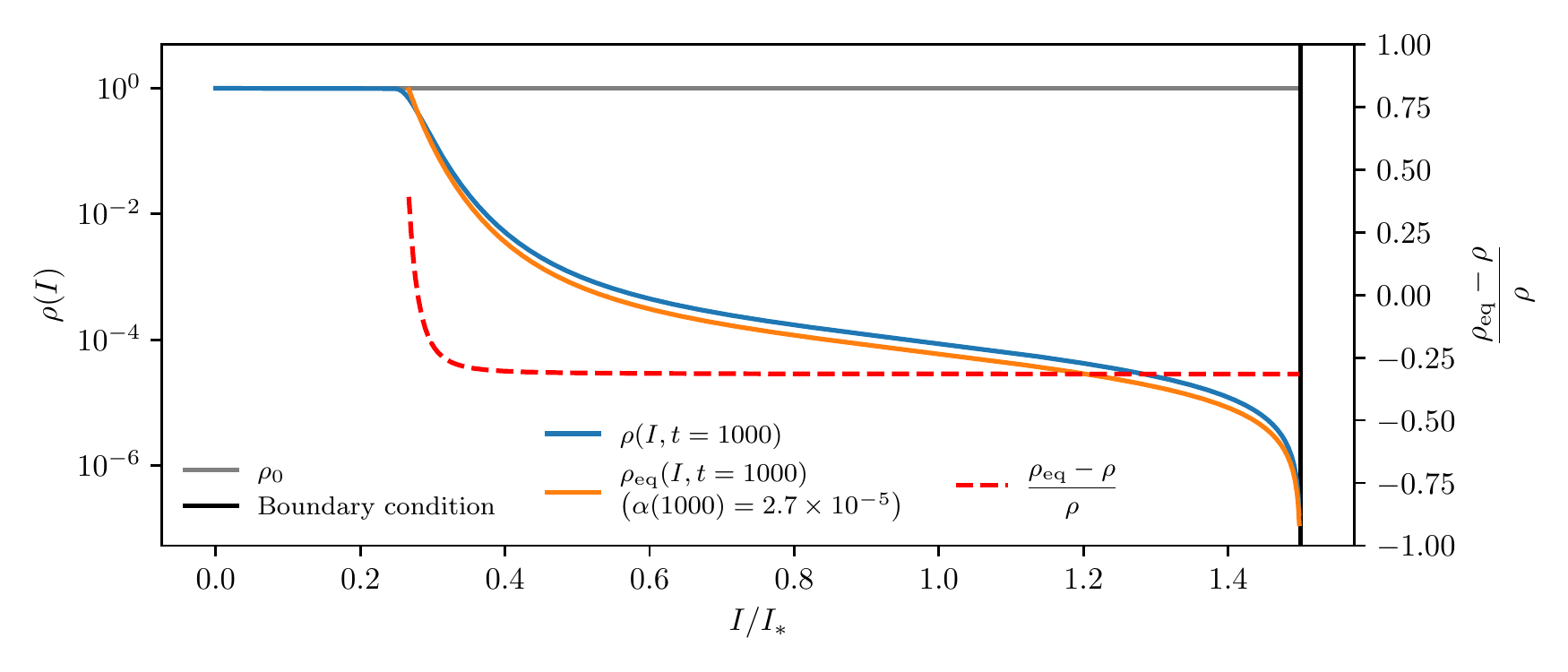}
    \caption{Initial uniform distribution for the simulation shown in Fig.~\ref{fig:1} after numerical integration at $t=1000 \, [\text{a. u.}]$, compared to the estimate of $\rho_\mathrm{eq}$ from Eq.~\eqref{eq:equilibrium_stationary_distribution}, with $\alpha(t)$ computed with Eq.~\eqref{eq:alpha_with_time}. (Simulations parameters: $(I_\ast = 1.0\,[\sigma^2],\, \kappa = 0.33)$).}
    \label{fig:3}
\end{figure*}

This behaviour can be observed in Fig.~\ref{fig:3}, where a Nekhoroshev-like diffusive process is simulated for a time long enough to reach the semi-stationary regime. Here, we consider the distribution obtained after prolonging the simulation of the system presented in Fig.~\ref{fig:1} with $\kappa=0.33$, and we compare it to $\rho_\mathrm{eq}$ from Eq.~\eqref{eq:equilibrium_stationary_distribution}, using $\alpha$ obtained from Eq.~\eqref{eq:alpha_with_time}. A global offset between the two curves is observed, which clearly highlights the limits of the approximation. 

%%%%%%%%%%%%%%%%%%%%%%%%%%%%%%%%%%%%%%%%%%%%%%%%%%%%%%%%%%%%%%%%%%%%%%%%%%%%%%%%

\section{Reconstruction of the diffusion coefficient of a FP process}
\label{sec:moving_the_absorbing_barrier}

%%%%%%%%%%%%%%%%%%%%%%%%%%%%%%%%%%%%%%%%%%%%%%%%%%%%%%%%%%%%%%%%%%%%%%%%%%%%%%%%

We consider now the problem of modelling the variation of the outgoing current after a change of the position $I_\mathrm{a}$ of the absorbing boundary condition, under the hypothesis that the movement is fast enough to be considered instantaneous and the movement is performed over a short distance while the system is in the semi-stationary regime described in the previous section. The ultimate goal consists in defining a method for probing the information about the shape of the diffusion coefficient $D(I)$ contained in the outgoing current measured after the instantaneous movement of the boundary condition. This, in view of reconstructing the characteristics of the FP process under analysis, which corresponds to evaluating the values of the two parameters $I_\ast$ and $\kappa$ defining $D(I)$.

We define two types of outgoing current, namely \textsl{global current}, i.e.\ the outgoing current observed from a slow core erosion process while keeping the absorbing boundary condition fixed, and \textsl{recovery current}, i.e.\ the current observed after the absorbing boundary condition is instantaneously moved and the system relaxes to a new semi-stationary regime.

We start by modelling the shape of the recovery current for a stationary system with a fixed source, both for inward and outward movements of the absorbing boundary. We recall that this mimics what is done experimentally when attempting to measure the diffusion equation by performing scans of the position of some collimators jaws~\cite{PhysRevSTAB.16.021003,PhysRevAccelBeams.23.044802}. Afterwards, we try to adapt the models to a system in semi-stationary regime, characterised by a source evolving with time.

%%%%%%%%%%%%%%%%%%%%%%%%%%%%%%%%%%%%%%%%%%%%%%%%%%%%%%%%%%%%%%%%%%%%%%%%%%%%%%%%

\subsection{Moving the absorbing boundary condition inwards}

%%%%%%%%%%%%%%%%%%%%%%%%%%%%%%%%%%%%%%%%%%%%%%%%%%%%%%%%%%%%%%%%%%%%%%%%%%%%%%%%

\begin{figure*}[htp]
    \centering
    \includegraphics{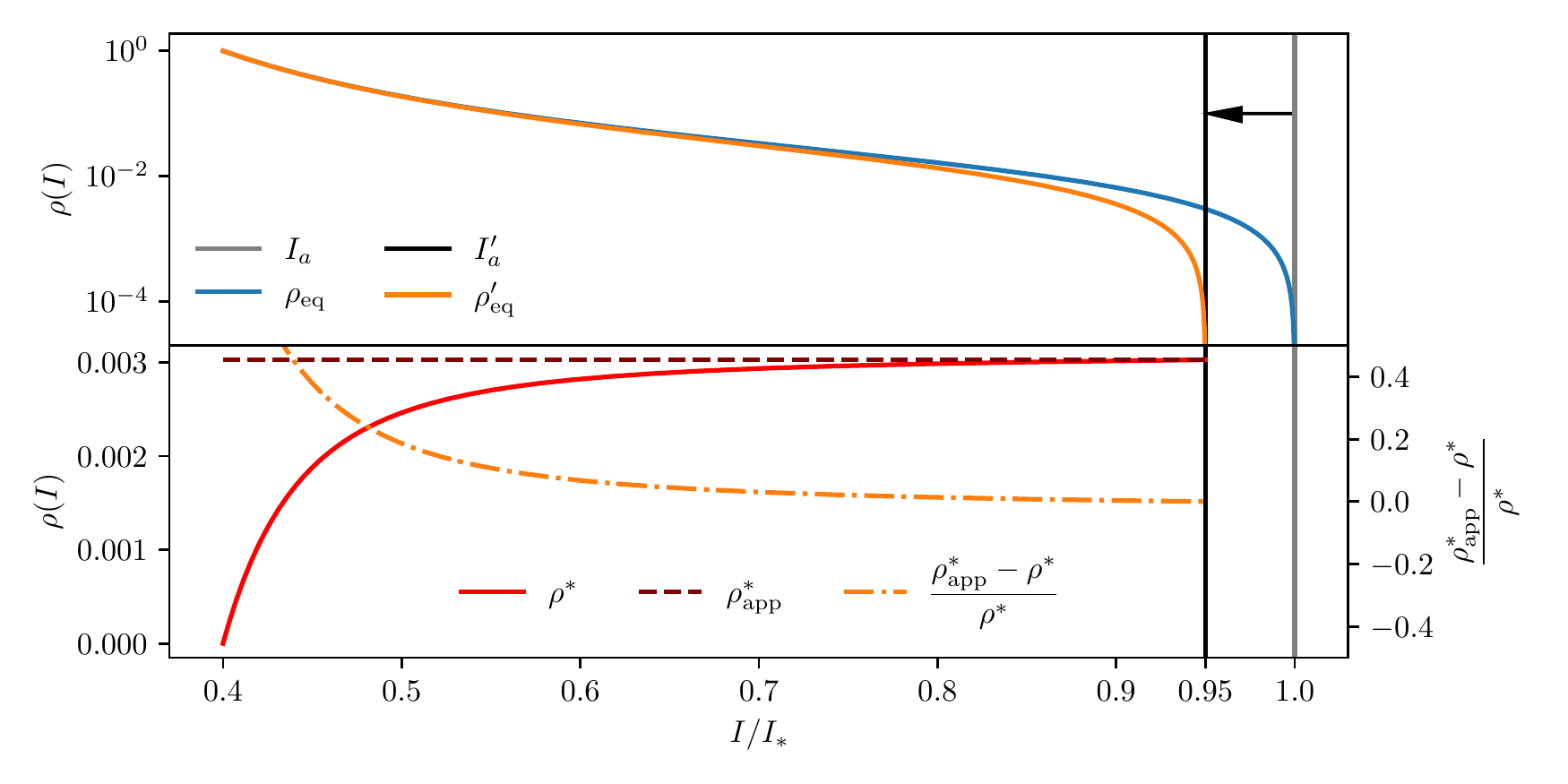}
    \caption{Top: Equilibrium distribution for $I_\mathrm{a}/I_\ast = 1.0$ compared with the  equilibrium distribution for $I'_\mathrm{a}/I_\ast = 0.95$. Bottom: Difference between the two equilibrium distributions. (Simulations parameters: $I_\ast = 1.0\,[\sigma^2],\, \kappa = 0.33,\, I_0/I_\ast = 0.4$).}
    \label{fig:4}
\end{figure*}

Let us consider a system in equilibrium with a constant  source $\rho(I_0, t)=1$, and an absorbing boundary condition $\rho(I_\mathrm{a}, t)=0$, and assume that the boundary condition is instantaneously moved inwards to $\rho(I'_\mathrm{a}, t)=0$, with $I_0 < I'_\mathrm{a} < I_\mathrm{a}$. After this change, the equilibrium distribution varies and the new equilibrium distribution is given by
\begin{equation}
    \rho'_\text{eq}(I) = \beta \int_I^{I'_\mathrm{a}} \frac{1}{D(x)}\,\mathrm{d}x\,,
\end{equation}
where $\beta$ is a constant such  that $\rho'_\text{eq}(I_0)=1$, and when compared with the constant $\alpha$ from Eq.~\eqref{eq:equilibrium_stationary_distribution}, we have that $\alpha < \beta$. The graphs of $\rho_\mathrm{eq}$ and $\rho'_\mathrm{eq}$ are shown in Fig.~\ref{fig:4} (top).

To apply the analytical formulae presented in the previous section, we need to compute the difference distribution $\rho^\ast(I)$. Assuming that the original system starts from the equilibrium distribution in Eq.~\eqref{eq:equilibrium_stationary_distribution}, we obtain 
\begin{align}
    \rho^\ast(I) = \rho_\text{eq}(I) - \rho'_\text{eq}(I) 
    &= \alpha \int_I^{I_\mathrm{a}} \frac{1}{D(x)}\,\mathrm{d}x - \beta \int_I^{I'_\mathrm{a}} \frac{1}{D(x)}\,\mathrm{d}x \notag \\
    &= \alpha \left( 
          \int_{I}^{I'_\mathrm{a}} \frac{1}{D(x)}\,\mathrm{d}x 
        + \int_{I'_\mathrm{a}}^{I_\mathrm{a}} \frac{1}{D(x)}\,\mathrm{d}x \right) - \beta \int_I^{I'_\mathrm{a}} \frac{1}{D(x)}\,\mathrm{d}x \notag \\
    &=  \alpha \int_{I'_\mathrm{a}}^{I_\mathrm{a}} \frac{1}{D(x)}\,\mathrm{d}x \, - (\beta - \alpha) \int_{I}^{I'_\mathrm{a}} \frac{1}{D(x)}\,\mathrm{d}x \notag \\
    &=  \rho^\ast_\text{app} - (\beta - \alpha) \int_{I}^{I'_\mathrm{a}} \frac{1}{D(x)}\,\mathrm{d}x \, .
    \label{eq:inward_difference}
\end{align}

The shape of $\rho^\ast(I)$ is shown in Fig.~\ref{fig:4} (bottom). This function, restricted to the interval $I \in [I_0, I'_\mathrm{a}]$, is monotonously increasing, with the maximum in $I'_\mathrm{a}$. Given the  Nekhoroshev-like form of the diffusion coefficient, the decrease to zero when $I \to I_0$ is exponentially fast, while in the region close to $I'_\mathrm{a}$, the function remains almost constant to the value  $\rho^\ast_\text{app}$ that can be used as the lowest-order approximation of $\rho^\ast(I)$.

\begin{figure*}[htp]
    \centering
    \includegraphics{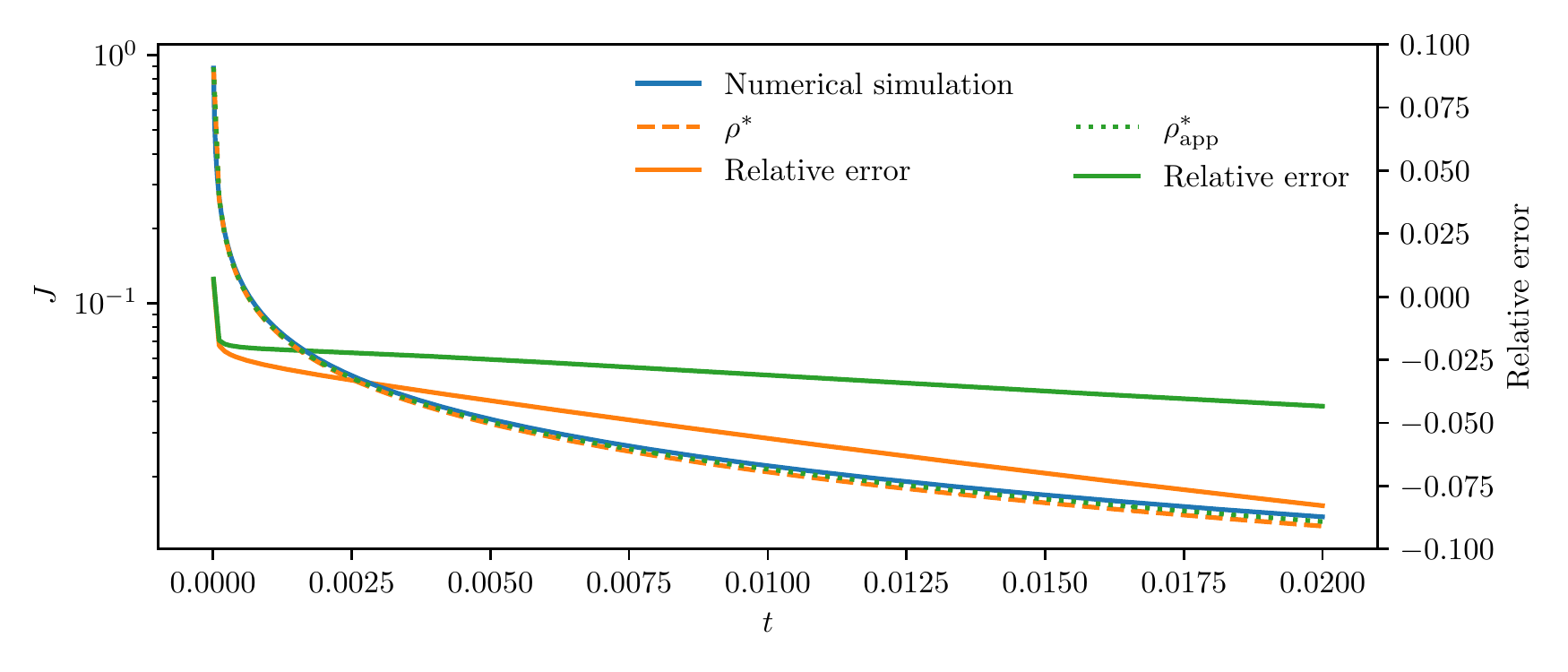}
    \caption{Comparison between the current generated by the inward displacement of the absorbing boundary condition shown in Fig.~\ref{fig:4}, its analytical estimate based on $\rho^\ast(I)$, and the analytical estimate based on $\rho^\ast_\text{app}$.}
    \label{fig:5}
\end{figure*}

In Fig.~\ref{fig:5}, we compare the simulated current with its analytical estimate, obtained by computing the convolution with the distribution $\rho^\ast(I)$ of Eq.~\eqref{eq:inward_difference}, which are in very good agreement. In the same figure, the analytical approximation based on the convolution with $\rho^\ast 
_\text{app}$ is shown, and even in this case the agreement is very good.

%%%%%%%%%%%%%%%%%%%%%%%%%%%%%%%%%%%%%%%%%%%%%%%%%%%%%%%%%%%%%%%%%%%%%%%%%%%%%%%%

\subsection{Moving the absorbing boundary condition outwards}

%%%%%%%%%%%%%%%%%%%%%%%%%%%%%%%%%%%%%%%%%%%%%%%%%%%%%%%%%%%%%%%%%%%%%%%%%%%%%%%%

\begin{figure*}[htp]
    \centering
    \includegraphics{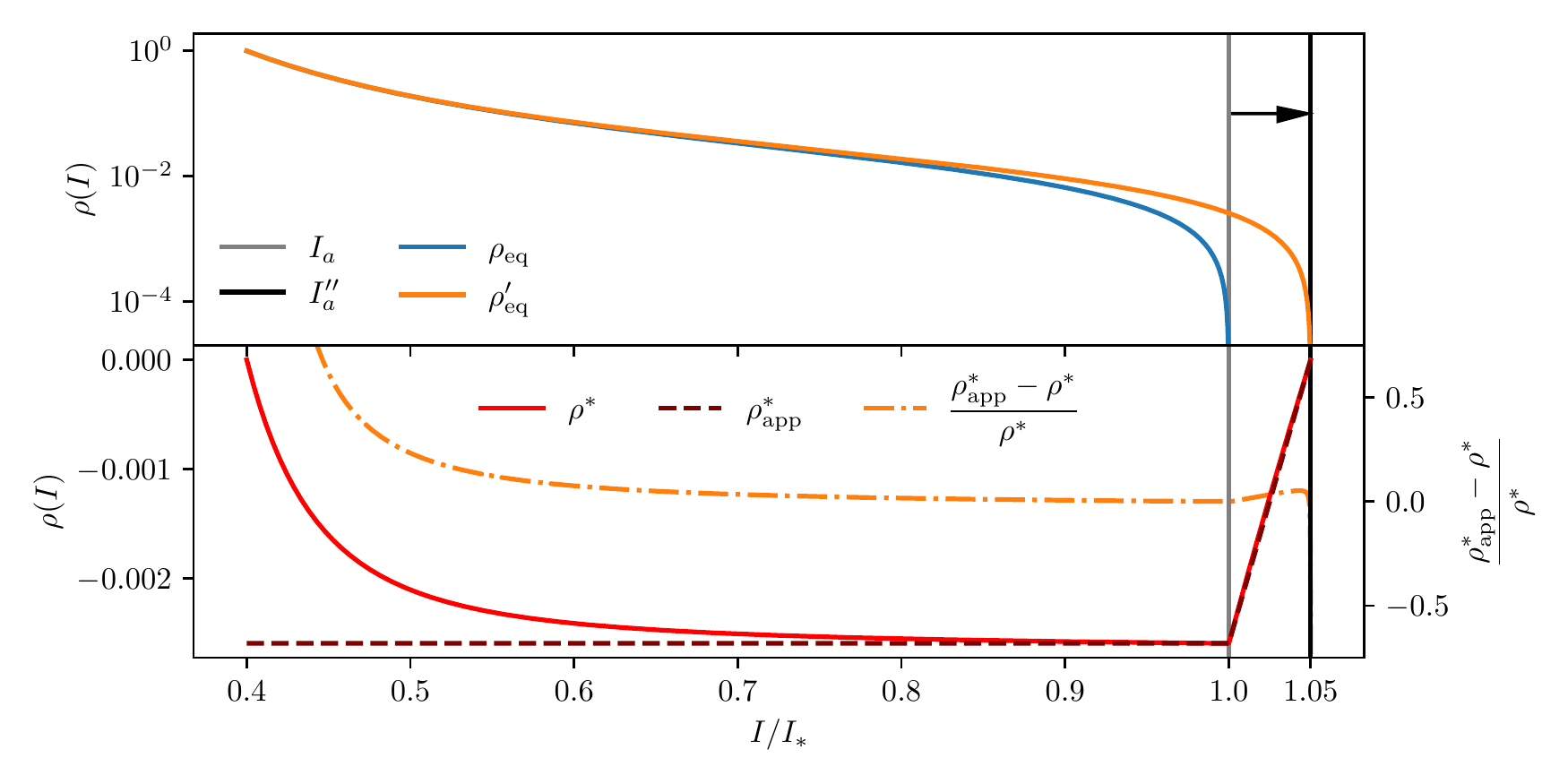}
    \caption{Top: Equilibrium distribution for $I_\mathrm{a}/I_\ast = 1.0$ compared with the equilibrium distribution for $I''_\mathrm{a}/I_\ast = 1.05$. Bottom: Difference between the two distributions. (Simulations parameters: $I_\ast = 1.0\,[\sigma^2],\, \kappa = 0.33,\, I_0/I_\ast = 0.4$).}
    \label{fig:6}
\end{figure*}

Let us now consider a system in equilibrium with a constant source $\rho(I_0, t)=1$, and an absorbing boundary condition $\rho(I_\mathrm{a}, t)=0$, and assume that this boundary condition is instantaneously moved outwards to $\rho(I''_\mathrm{a}, t)=0$, with $I_0 < I_\mathrm{a} < I''_\mathrm{a}$. The new equilibrium distribution is given by
\begin{equation}
    \rho''_\text{eq}(I) = \gamma \int_I^{I''_\mathrm{a}} \frac{1}{D(x)}\,\mathrm{d}x\,,
\end{equation}
where $\gamma$ is a constant such that $\rho''_\text{eq}(I_0)=1$, and when compared with the constant $\alpha$ from Eq.~\eqref{eq:equilibrium_stationary_distribution}, we have that $\gamma < \alpha$. The graphs of $\rho_\text{eq}(I)$ and $\rho''_\text{eq}(I)$ are shown in Fig.~\ref{fig:6} (top). 

To define properly the difference distribution in the new interval $[I_0, I''_\mathrm{a}]$, we need to extend the definition of the equilibrium distribution $\rho_\text{eq}(I)$, namely
\begin{equation}
    \rho_\text{eq}(I) = 
    \left\{\begin{array}{lr}
        \alpha \displaystyle{\int_I^{I_\mathrm{a}} \frac{1}{D(x)}\,\mathrm{d}x} \qquad & \text{if } \, I \leq I_\mathrm{a} \\
        \\
        0 \qquad & \text{if } \, I > I_\mathrm{a}\,,
    \end{array}\right.
\end{equation}
which leads to the following expression for the difference distribution  
\begin{equation}
    \rho^\ast(I) = 
    \left\{\begin{array}{lr}
        - \gamma \displaystyle{\int_{I_\mathrm{a}}^{I''_\mathrm{a}} \frac{1}{D(x)}\,\mathrm{d}x + (\alpha - \gamma) \int_{I}^{I_\mathrm{a}} \frac{1}{D(x)}\,\mathrm{d}x}\ \quad &\text{if} \, I \leq I_\mathrm{a}\\
        \\
        - \gamma \displaystyle{\int_{I}^{I''_\mathrm{a}} \frac{1}{D(x)}\,\mathrm{d}x} \quad &\text{if} \, I > I_\mathrm{a}\,,
    \end{array}\right. 
    \label{eq:outward_difference}
\end{equation}
which is a negative distribution, with a minimum at $I_\mathrm{a}$ and with $\rho^\ast(I_0) = \rho^\ast(I''_\mathrm{a}) = 0$.

A plot of $\rho^\ast(I)$ is shown in Fig.~\ref{fig:6} (bottom), and we remark that this distribution leads to a negative outgoing current that needs to be combined with the stationary current from the equilibrium process for obtaining the actual outgoing current.

While in the interval $[I_0, I_\mathrm{a}]$ a constant approximated distribution function can be a reasonable assumption, in the interval $[I_\mathrm{a}, I''_\mathrm{a}]$ a different approximation is needed. Under the assumption that the outward step $I''_\mathrm{a} - I_\mathrm{a}$ is small, a linear approximation from $\rho^\ast(I_\mathrm{a})$ to $\rho^\ast(I''_\mathrm{a})$ can be considered, namely
\begin{equation}
    \rho^\ast_\text{app}(I) = 
    \left\{\begin{array}{lr}
        - \gamma \displaystyle{\int_{I_\mathrm{a}}^{I''_\mathrm{a}} \frac{1}{D(x)}\,\mathrm{d}x} \quad &\text{  if } \, I \leq I_\mathrm{a}\\
        \\
        - \gamma \displaystyle{\left(\frac{I''_\mathrm{a} - I}{I''_\mathrm{a} - I_\mathrm{a}} \right)} \displaystyle{\int_{I_\mathrm{a}}^{I''_\mathrm{a}} \frac{1}{D(x)}\,\mathrm{d}x} \quad &\text{  if } \, I > I_\mathrm{a} \,.
    \end{array}\right. 
    \label{eq:outward_difference_approx}
\end{equation} 

\begin{figure*}[htp]
    \centering
    \includegraphics{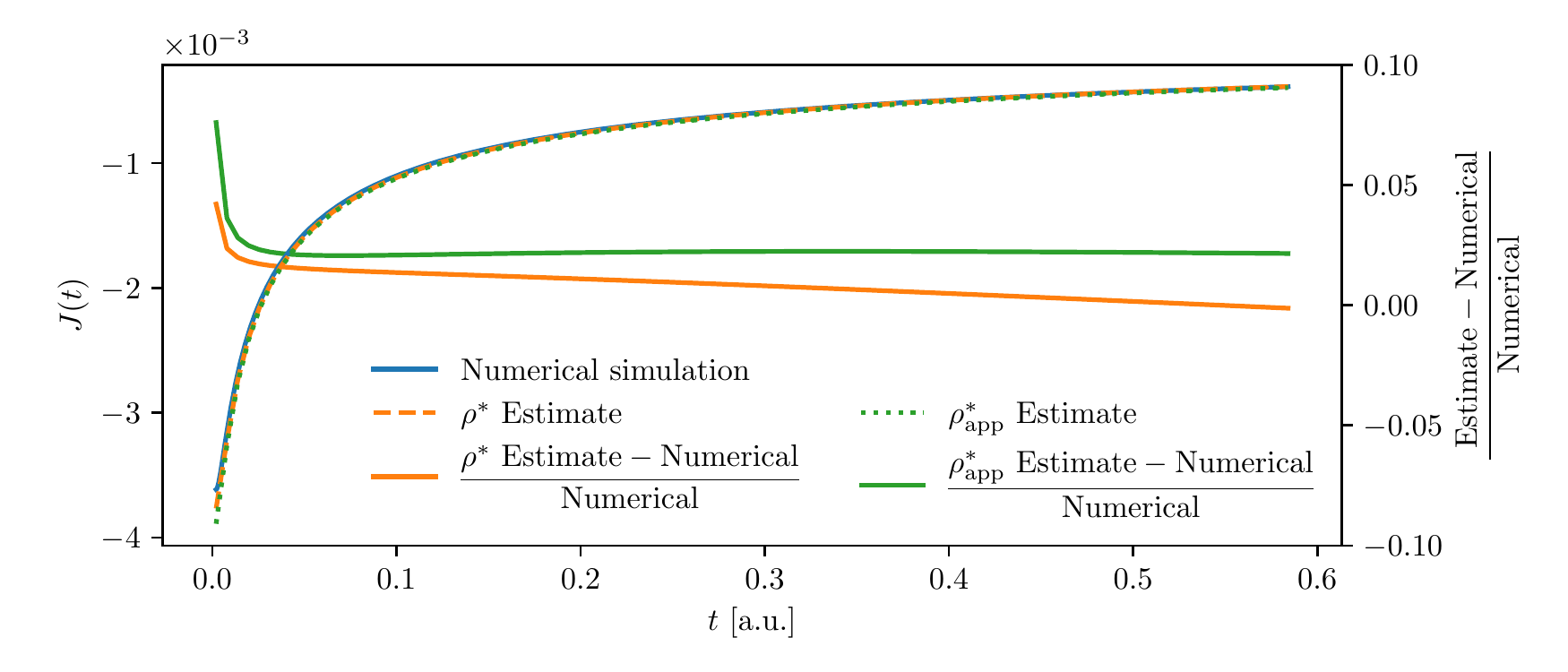}
    \caption{Comparison between the current generated by the outward displacement of the absorbing boundary condition shown in Fig.~\ref{fig:6}, its analytical estimate based on $\rho^\ast(I)$, and the analytical estimate based on the approximation $\rho^\ast_\text{app}$.}
    \label{fig:7}
\end{figure*}

In Fig.~\ref{fig:7}, we compare the simulated current with its analytical estimate, obtained by computing the convolution with the distribution $\rho^\ast(I)$ of Eq.~\eqref{eq:outward_difference}, which are in very good agreement. In the same figure, the analytical approximation based on the convolution with $\rho^\ast 
_\text{app}$ is shown, and even in this case the agreement is very good.

%%%%%%%%%%%%%%%%%%%%%%%%%%%%%%%%%%%%%%%%%%%%%%%%%%%%%%%%%%%%%%%%%%%%%%%%%%%%%%%%

\subsection{Moving the absorbing boundary condition in a semi-stationary system}
\label{subsec:Moving_the_boundary_in_a_semi_stationary_system}

%%%%%%%%%%%%%%%%%%%%%%%%%%%%%%%%%%%%%%%%%%%%%%%%%%%%%%%%%%%%%%%%%%%%%%%%%%%%%%%%

In Section~\ref{subsec:semi_stationary_regime_for_a_real_system} we have seen how it is possible to describe a diffusive process, after a transient time, as a semi-stationary process in which a stable-core is being slowly eroded over an exponentially-long time, with an approximated timescale given by Eq.~\eqref{eq:peak_current_time}. If the position of the absorbing boundary condition is changed when the system is in this semi-stationary state, and the new position is close to the original one, so that it is characterised by a timescale of the same order of magnitude, the stationary part of the system, i.e.\ the relaxed part outside of the stable-core, will relax to a new configuration in a time that is short compared to the time scale of the stable-core erosion.

Being the timescale of the recovery-current process orders of magnitude shorter than the evolution of the global current, the variation of the shape of the core is so slow that it can be neglected. Hence, one can treat this situation as a source at a fixed position with a slow-varying intensity $\alpha(t)$. 

%%%%%%%%%%%%%%%%%%%%%%%%%%%%%%%%%%%%%%%%%%%%%%%%%%%%%%%%%%%%%%%%%%%%%%%%%%%%%%%%

\subsubsection{Normalising a recovery current}

%%%%%%%%%%%%%%%%%%%%%%%%%%%%%%%%%%%%%%%%%%%%%%%%%%%%%%%%%%%%%%%%%%%%%%%%%%%%%%%%

Thanks to the previous assumptions, one can define a normalisation procedure to be applied to the recovery current to make it independent of the characteristics of the global current. We are interested in reducing the problem to the ideal case of a constant source at $I_0$ and a constant unitary outgoing current at the absorbing boundary condition $I_\mathrm{a}$, instead of a system with a slow-varying global current $\alpha(t)$. This approach is tested by simulating the same Nekhoroshev-like FP system twice: firstly, by keeping the absorbing boundary fixed; secondly, by executing some instantaneous changes of the absorbing boundary position. With this approach, we gather information on the value of the semi-stationary current $\alpha(t)$, thus enabling the transformation of the process with a moving boundary to a system with a fixed source.

We define the normalised recovery current as the current obtained in the measurement with the moving absorbing barrier divided by the current $\alpha(t)$, obtained in the measurement with fixed boundary. The normalised recovery current has a unitary value when the absorbing boundary condition is not changed, and has normalised maximum and minimum, respectively, for the inward and outward movements of the absorbing boundary condition. The behaviour of the normalised recovery current can be related to the ideal stationary systems described above, for which analytical approximations are known.

\begin{figure*}[htp]
    \centering
    \includegraphics{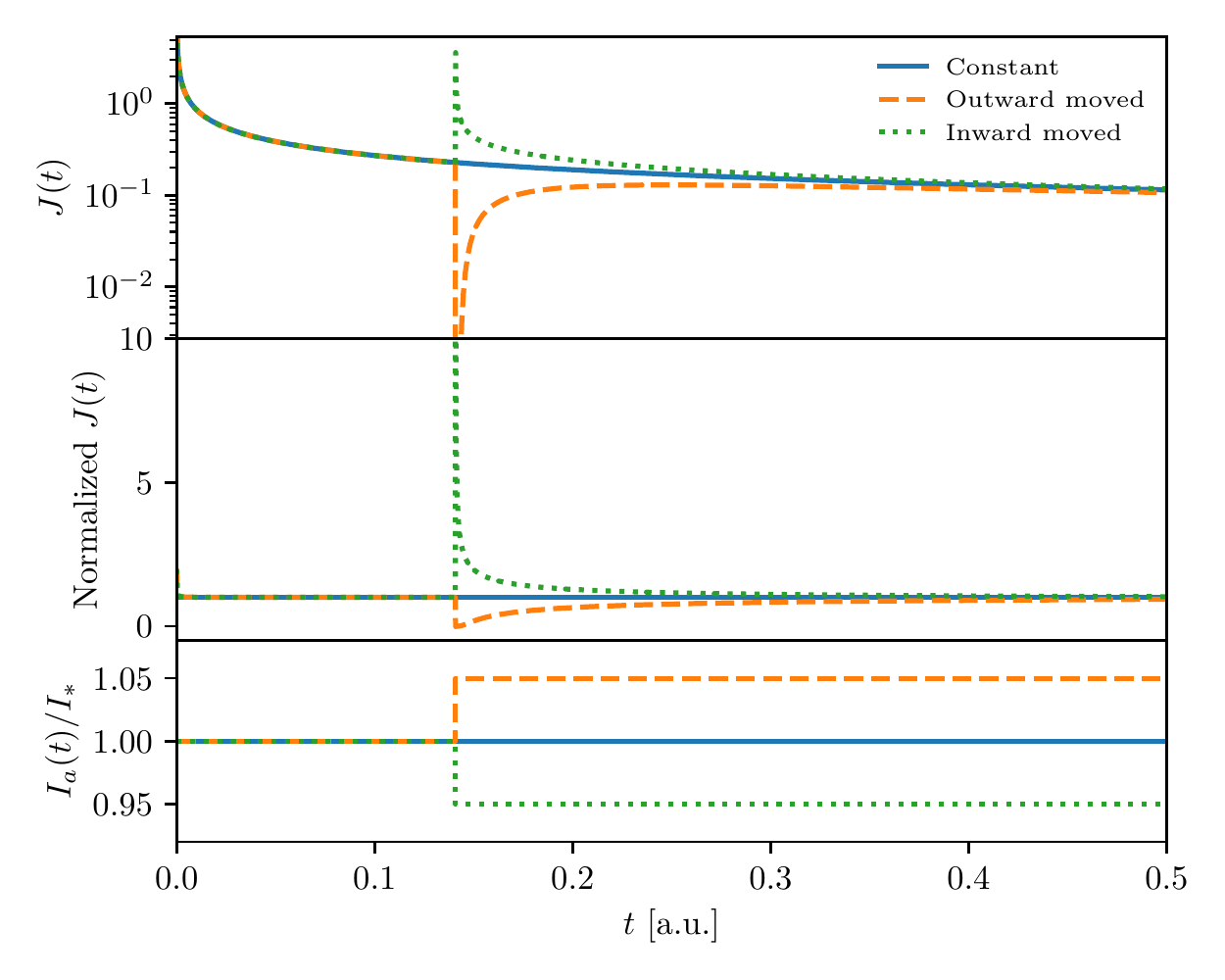}
    \caption{Top: Evolution of the outgoing current for three diffusive processes with different boundary conditions. Middle: Evolution of the normalised outgoing current. Bottom: Changes to the absorbing boundary condition in the three scenarios. (Simulations parameters: $I_\ast = 1.0\,[\sigma^2], \, \kappa = 0.33$).}
    \label{fig:fixed-vs-moved-boundary}
\end{figure*}

In Fig.~\ref{fig:fixed-vs-moved-boundary}, an example of such a procedure is displayed (centre) together with the evolution of the outgoing current (top) and the corresponding variation of the position of the boundary condition (bottom). It is worth mentioning that for every change of $I_\mathrm{a}$ the value of $\alpha(t)$ changes according to Eq.~\eqref{eq:alpha_with_time}, although for small variations of the absorbing boundary condition, a Taylor expansion can be applied.

%%%%%%%%%%%%%%%%%%%%%%%%%%%%%%%%%%%%%%%%%%%%%%%%%%%%%%%%%%%%%%%%%%%%%%%%%%%%%%%%

\subsubsection{Normalising a recovery current without knowledge of the global current}

%%%%%%%%%%%%%%%%%%%%%%%%%%%%%%%%%%%%%%%%%%%%%%%%%%%%%%%%%%%%%%%%%%%%%%%%%%%%%%%%

Whenever it is not possible to repeat two complete measurements on the same process, i.e.\ one with and one without changes of the boundary condition position, the normalisation procedure defined previously needs to be adapted.

Ideally, the best strategy consists in waiting long enough after each change of the position of the absorbing boundary to reach an equilibrium and to accomplish the recovery process so that the outgoing current measured before the change of boundary position and after the long wait is a pure global current. This would approximately correspond to a complete relaxation of the difference distribution resulting from the boundary movement. Hence, in this way, the outgoing current can be used for reconstructing the shape of the global current and for the normalisation procedure.

However, some extra hurdles should be considered: since we do not have a prior knowledge of the value of $D(I)$, we do not know the timescales of the recovery currents or those of the core-eroding process. Moreover, even though a good fraction of the recover process is achieved very quickly, a full recover, corresponding to a full relaxation of the difference distribution $\rho^\ast$, might take an exponentially-long amount of time, possibly beyond computing capabilities. Hence, it might not be possible to perform such a long measurement in a particle accelerator. Therefore, it is necessary to define a protocol that enables quantitative criteria to establish whether or not an assumed recovery time is long enough to ensure a meaningful reconstruction of the behaviour of the FP process, possibly including an estimate of the uncertainty in the reconstruction of the global current.

\begin{figure*}[t]
    \centering
    \includegraphics{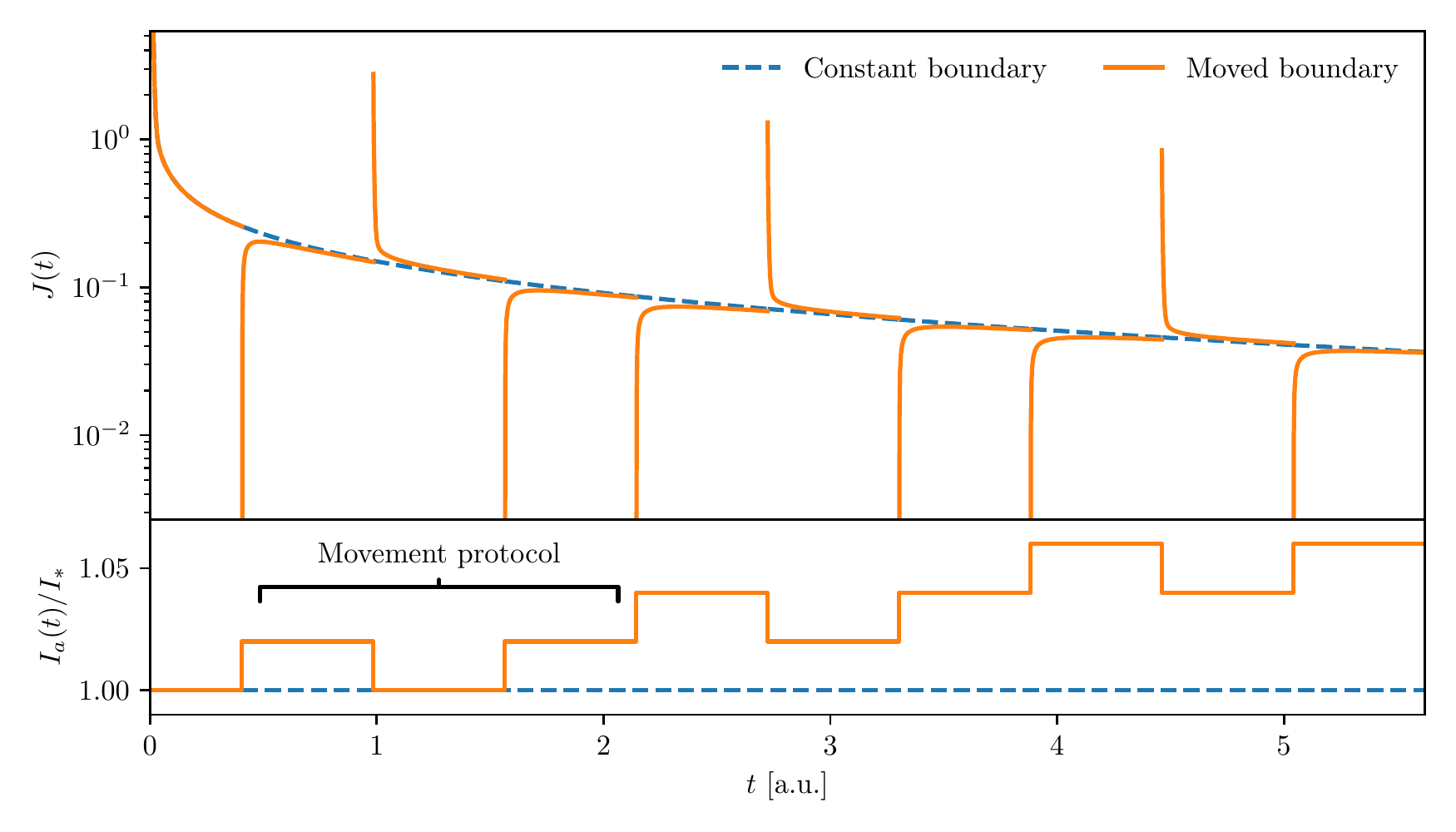}
    \caption{Example of the proposed three-step protocol with a direct comparison between the condition with and without the variations in the position of the absorbing boundary condition. In this figure, the protocol is executed three times. The top plot shows the evolution of the outgoing current, while the bottom one shows the corresponding evolution of the position of the absorbing boundary condition. Note how the repetition of the three-step protocol moves progressively the absorbing boundary outwards. (Simulations parameters: $I_\ast = 1.0\,[\sigma^2],\,\kappa = 0.33$).}
    \label{fig:9}
\end{figure*}

A possible solution, compatible with these constraints, consists in the combination of three movements of the boundary condition, where, for each value of the action to be probed, an outward-inward-outward sequence of movements is performed. These three steps must be performed with a fixed movement size $\Delta I$ and with a fixed relaxation time $\Delta t$ between a boundary condition movement and the next one. The optimal values for $\Delta I$ and $\Delta t$ are discussed in the next section. A visualisation of this protocol is provided in Fig.~\ref{fig:9} (bottom), where the three-step boundary condition change is highlighted and repeated three times, and the corresponding evolution of the outgoing current is given (top), as computed from numerical simulations. As a comparison, the situation corresponding to the constant position of the boundary condition is also shown.

We remark that variants of the proposed three-step movement are indeed possible, and that this proposal is also motivated by the wish not to introduce unnecessary complications. It is also important stressing that the assumption of performing small and equal movements of the absorbing barrier position for the three steps implies that also $\Delta t$ should be the same for the steps, as the relaxation time should be approximately the same for all steps. 

This three-step sequence of absorbing-barrier changes is performed at different values of the global current, and this basic sequence can be repeated by performing it at different action values. The resulting sequence of alternating recovery currents provides an approximation of the evolution of the global current with a sequence of upper- and lower-bound values at different times, which can be interpolated and used for the construction of a global current estimate. These bounds provide a degree of uncertainty directly linked to the chosen value $\Delta t$, as the longer is $\Delta t$ the lower is the degree of uncertainty in the reconstruction of the global current. A more detailed discussion on a possible quantitative definition of the optimal choice of the relaxation timescale $\Delta t$, together with the effects of using shorter relaxation times, is discussed in the next section.

\begin{figure*}[!ht]
    \centering    
    \includegraphics{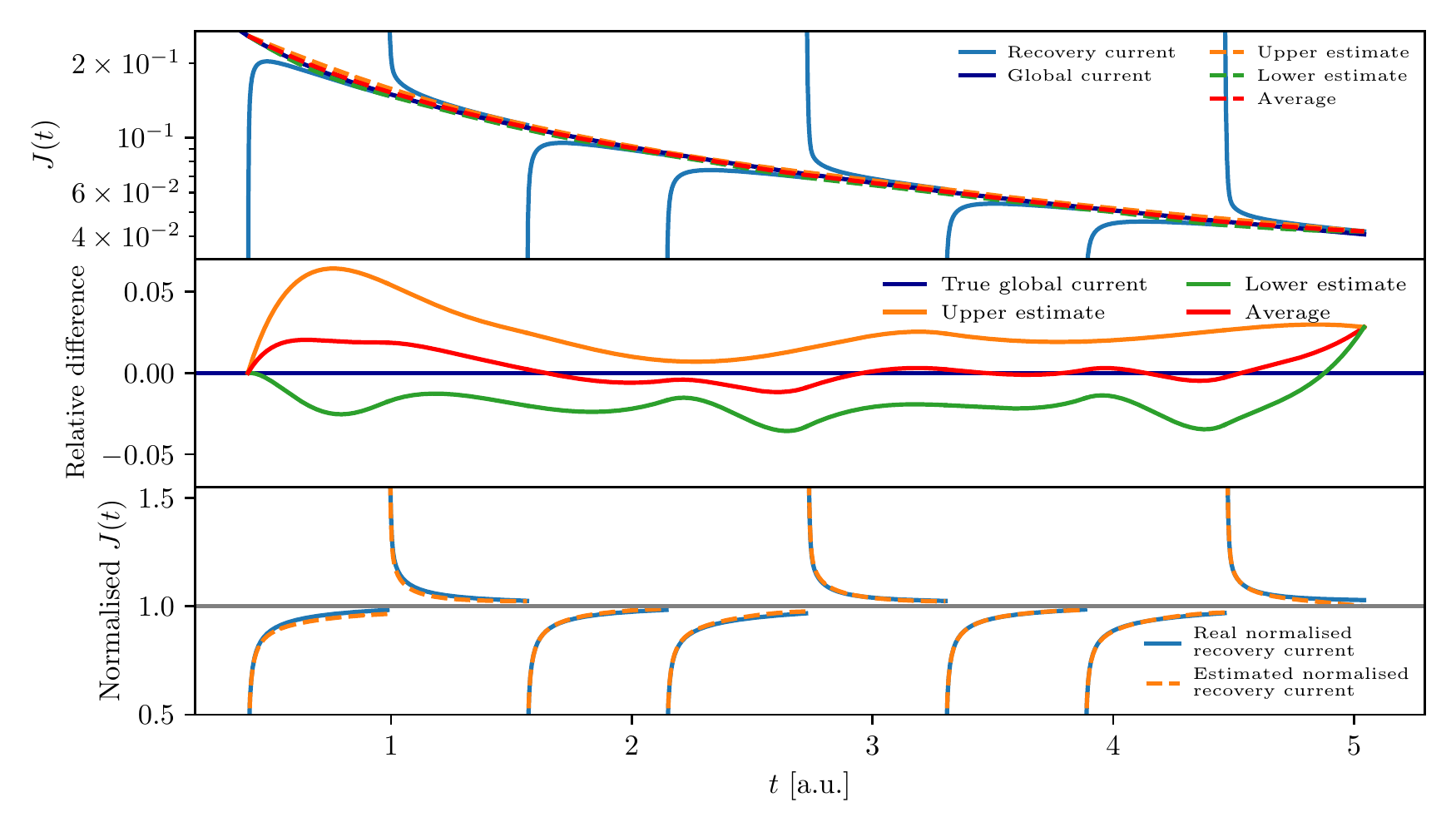}
    \caption{Example of the reconstruction process of the global current applied to a fraction of the data shown in Fig.~\ref{fig:9}. Top: Outgoing current from the process with constant boundary (dark blue) and with varying boundary conditions (light blue), along with the three components of the interpolation process. Middle: Relative difference, between the lower, average, or upper estimate of the global current and the true global current, of the interpolation procedure. Bottom: Comparison between the actual recovery current and the reconstructed one.}
    \label{fig:protocol}
\end{figure*}

To reconstruct the global current, an upper- and lower-bound estimate are derived by considering the last values of the inward and outward recovery currents, respectively. Two extra points are added to the upper- and lower-bound estimate, with the goal of covering the maximum time span for reconstructing the global current: the last measured global current value before the first boundary movement is added to both estimates; the last value of the last recovery current measured, which, being an outward recovery current, was already part of the lower-bound estimate, is added also to the upper-bound estimate. This explains why in Fig.~\ref{fig:protocol} (centre) the estimates coincide at the beginning and end of the interpolation interval. The two sets of upper-bound and lower-bound points are each interpolated with an univariate cubic spline and the average function of these two interpolating functions is taken as the estimate of the global current. 

We remark that the univariate cubic spline is taken with a number of knots so that the second derivative does not change in sign. This ensures that the resulting global current estimate fulfils the expected features of the actual global current. In particular, it avoids local oscillations that might be generated by a simple interpolation of the upper-bound and lower-bound points. The result of such an approach is shown in Fig.~\ref{fig:protocol} where a fraction of the data presented in Fig.~\ref{fig:9} is used for reconstructing the global current, and a good overall agreement with the actual global current is clearly visible.

%%%%%%%%%%%%%%%%%%%%%%%%%%%%%%%%%%%%%%%%%%%%%%%%%%%%%%%%%%%%%%%%%%%%%%%%%%%%%%%

\subsection{Reconstructing $D(I)$ from the normalised recovery currents}

%%%%%%%%%%%%%%%%%%%%%%%%%%%%%%%%%%%%%%%%%%%%%%%%%%%%%%%%%%%%%%%%%%%%%%%%%%%%%%%

After performing the proposed reconstruction protocol, a series of normalised recovery currents, obtained for different positions of the boundary condition, are available. All these curves are then used to reconstruct the shape of $D(I)$ via a fit procedure.

Thanks to the normalisation procedure, every normalised recovery current can be considered as an individual and independent relaxation process, as in a system in equilibrium with a fixed source. Therefore, we can consider as expected current the convolution of the analytical current, presented in Eq.~\eqref{eq:thecurrent}, with one of our approximated difference distribution $\rho^\ast$. From a normalised recovery current obtained from an inward movement, we expect a relaxation curve characterised by an equivalent process with an initial distribution given by Eq.~\eqref{eq:inward_difference}, where the value of $\rho^\ast_\text{app}$ is computed considering $\alpha=1$, due to the normalisation performed, the integral being computed over the appropriate action interval. Likewise, for a recovery current from an outward movement, we expect a curve characterised by an equivalent process with an initial distribution given by Eq.~\eqref{eq:outward_difference_approx}, where we consider $\gamma=1$, due to the normalisation performed, the integral being computed over the appropriate action interval. Assuming a Nekhoroshev-like form for $D(I)$, the goal is to determine the values of the two parameters $I_\ast$ and $\kappa$, which is obtained by a standard non-linear least squares algorithm applied to the currents obtained during the execution of the proposed protocol.

%%%%%%%%%%%%%%%%%%%%%%%%%%%%%%%%%%%%%%%%%%%%%%%%%%%%%%%%%%%%%%%%%%%%%%%%%%%%%%%%

\section{Numerical results}
\label{sec:numerical_results}

%%%%%%%%%%%%%%%%%%%%%%%%%%%%%%%%%%%%%%%%%%%%%%%%%%%%%%%%%%%%%%%%%%%%%%%%%%%%%%%%

To test the validity of the proposed procedure and to obtain a complete overview of its performance and limitations in the reconstruction of $D(I)$, several numerical simulations of diffusive processes with a Nekhoroshev-like diffusion coefficient have been performed by using the protocol described in Section~\ref{sec:moving_the_absorbing_barrier}. Particular emphasis is given to establishing the reliability of the proposed procedure as a function of the values of $\Delta I$ and $\Delta t$ used in the protocol of variation of the position of the boundary conditions.

%%%%%%%%%%%%%%%%%%%%%%%%%%%%%%%%%%%%%%%%%%%%%%%%%%%%%%%%%%%%%%%%%%%%%%%%%%%%%%%%

\subsection{Simulation parameters}

%%%%%%%%%%%%%%%%%%%%%%%%%%%%%%%%%%%%%%%%%%%%%%%%%%%%%%%%%%%%%%%%%%%%%%%%%%%%%%%%

As an initial condition, we consider the distribution in Eq~\eqref{eq:initial_distribution}, and we remark that all action variables are taken in units of sigma. We then consider a Nekhoroshev-like diffusive system characterised by the parameters obtained from the studies reported in Ref.~\cite{bazzani2020diffusion}, namely $I_\ast = 20.0\,[\sigma^2]$, $\kappa = 0.33$. Such a system, as it can be seen from Figs.~\ref{fig:1} and~\ref{fig:3}, is compatible with the semi-stationary regime and hence to the application of the proposed procedure.

Different values of the starting position $I_\mathrm{a}/I_\ast$ of the absorbing boundary condition have been considered. Even though most of our assumptions are valid for the $I_\mathrm{a}/I_\ast < 1$ regime, we consider also starting positions near $I_\ast$ and beyond $I_\ast$, i.e.\ in the saturation region of $D(I)$, to evaluate how robust the method is in non-ideal conditions.

For each configuration, after an initial time delay when a semi-stationary regime is reached, ten repetitions of the three-step protocol (outward-inward-outward), displayed in Fig.~\ref{fig:9}, have been performed. Several values of $\Delta I$, i.e.\ the action change of the boundary condition position, have been used to assess the presence of an optimal value for the reconstruction procedure. We remark that at the end of a simulation, the position of the absorbing boundary condition has moved from $I_\mathrm{a}/I_\ast$ to $(I_\mathrm{a} + 10\Delta I) / I_\ast$. 

Several values of the relaxation time $\Delta t$ have been considered to assess the reconstruction performance under different levels of equilibrium. We remark that an empirical relaxation time has been defined as the time for which a normalised recovery current is expected to recover the $99.9\%$ of the value of the original global current. Such an ideal time is computed using the full knowledge of $D(I)$, using our analytic current estimate Eq.~\eqref{eq:thecurrent}, and considering an outward movement of the absorbing boundary condition of size $\Delta I$ from the initial position of the absorbing boundary. It is stressed that, in general, we should assume that such relaxation time is not known when reconstructing the value of the diffusion coefficient. It is also worth mentioning that a criterion based on a complete $100\%$ recover of the global current cannot be used in practice, as this would require exponentially-long simulation times, needed to reach the relaxation of the inner part of the distribution, with negligible differences with respect to the $99.9\%$ case. Different fractions of this ideal time have been used when performing our procedure, and we evaluate how times shorter than the ideal relaxation time impact the quality of our final fit, as the system is still in a non-equilibrium regime when the next absorbing boundary movement occurs.

When working with the datasets generated by the various numerical simulations, a post-processing step on the normalised recovery currents is performed before executing the final fit procedure for reconstructing $D(I)$. It consists in selecting a fraction of the data representing the normalised recovery currents, i.e.\ only the normalised recovery current data up to a given percentage of the full recovery. For example, if we decide to filter out the normalised data beyond the $90\%$ recovery, it means that we discard values that are lower than $1.1$ for inward normalised recovery currents and values that are higher than $0.9$ for outward normalised recovering currents. We recall that, in the context of a normalised recovery current, a full recovery implies a value of $1.0$ as normalised recovery current.

\begin{figure*}[htp]
    \centering 
    \includegraphics[width=0.9\textwidth]{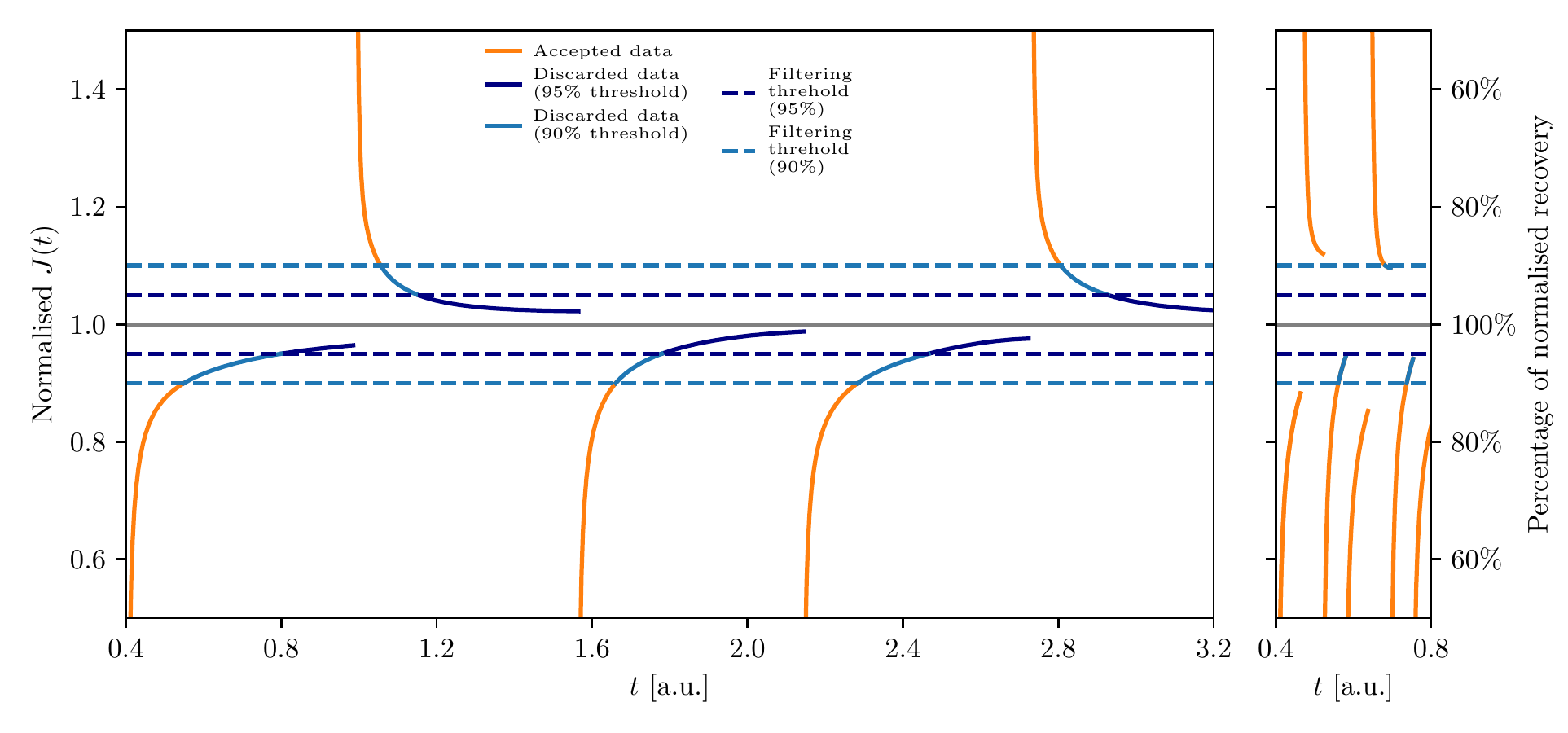}
    \caption{Left: normalised recovery current, shown already in Fig.~\ref{fig:protocol}, together with two filtering levels. The boundary movements are performed after an initial evolution time $t=0.4 \, [\text{a. u.}]$, and the relaxation time $\Delta t$ between one boundary movement and the next one is equal to $\Delta t=0.58 \, [\text{a. u.}]$. Right: the same system is simulated with the same initial evolution time $t=0.4 \, [\text{a. u.}]$ and one order of magnitude shorter relaxation time $\Delta t=0.058 \, [\text{a. u.}]$, the resulting normalised recovery currents have not relaxed long enough to reach the $95\%$ filtering level. When the selected filtering level is not reached by the normalised recovery currents, the whole dataset is used for the fit reconstruction and no parts are discarded.}
    \label{fig:postprocessing}
\end{figure*}

This post-processing step is displayed in Fig.~\ref{fig:postprocessing}, where two different values of the fraction of the relaxation time between boundary movements are used in the numerical simulations. In both simulations, the boundary movement starts after an equal waiting time. In the left plot, the normalised recovery currents, reported already in Fig.~\ref{fig:protocol}, are shown together  with two different filtering levels. In the right plot, the same system is simulated using a shorter fraction of the relaxation time, and the recovery currents are shown together with the same filtering levels presented in the left plot. The much shorter time leads to only a partial recovery of the currents between boundary condition changes. For these sets of normalised recovery currents, the filtering levels displayed lead to almost no data reduction.

By selecting different levels of filtering data, it is possible to evaluate how this choice affects the accuracy of the $D(I)$ reconstruction. It should be noted that our analytical approximation of Eq.~\eqref{eq:thecurrent} performs best when describing the evolution of a distribution near the absorbing boundary condition~\cite{montanari:ipac2021:tupab233}. Furthermore, the recovery current features an exponential-like decay that makes the analytical approximation less accurate over long time scales. For this reason, probing the dependence of the reconstruction performance on the fraction of data selected is very relevant.

%%%%%%%%%%%%%%%%%%%%%%%%%%%%%%%%%%%%%%%%%%%%%%%%%%%%%%%%%%%%%%%%%%%%%%%%%%%%%%%%

\subsection{Analysis of the reconstruction performance}

%%%%%%%%%%%%%%%%%%%%%%%%%%%%%%%%%%%%%%%%%%%%%%%%%%%%%%%%%%%%%%%%%%%%%%%%%%%%%%%%

The numerical exploration of the FP process entails a scan over several simulation parameters, leading to a large hyperspace of possible configurations. For this reason, we focus on the most ideal configurations, i.e.\ those that provide the best reconstruction performance, and then show how the other parameters affect the reconstruction accuracy of $I_\ast$ and $\kappa$.

\begin{figure*}[htp]
    \centering
    \includegraphics[width=\textwidth]{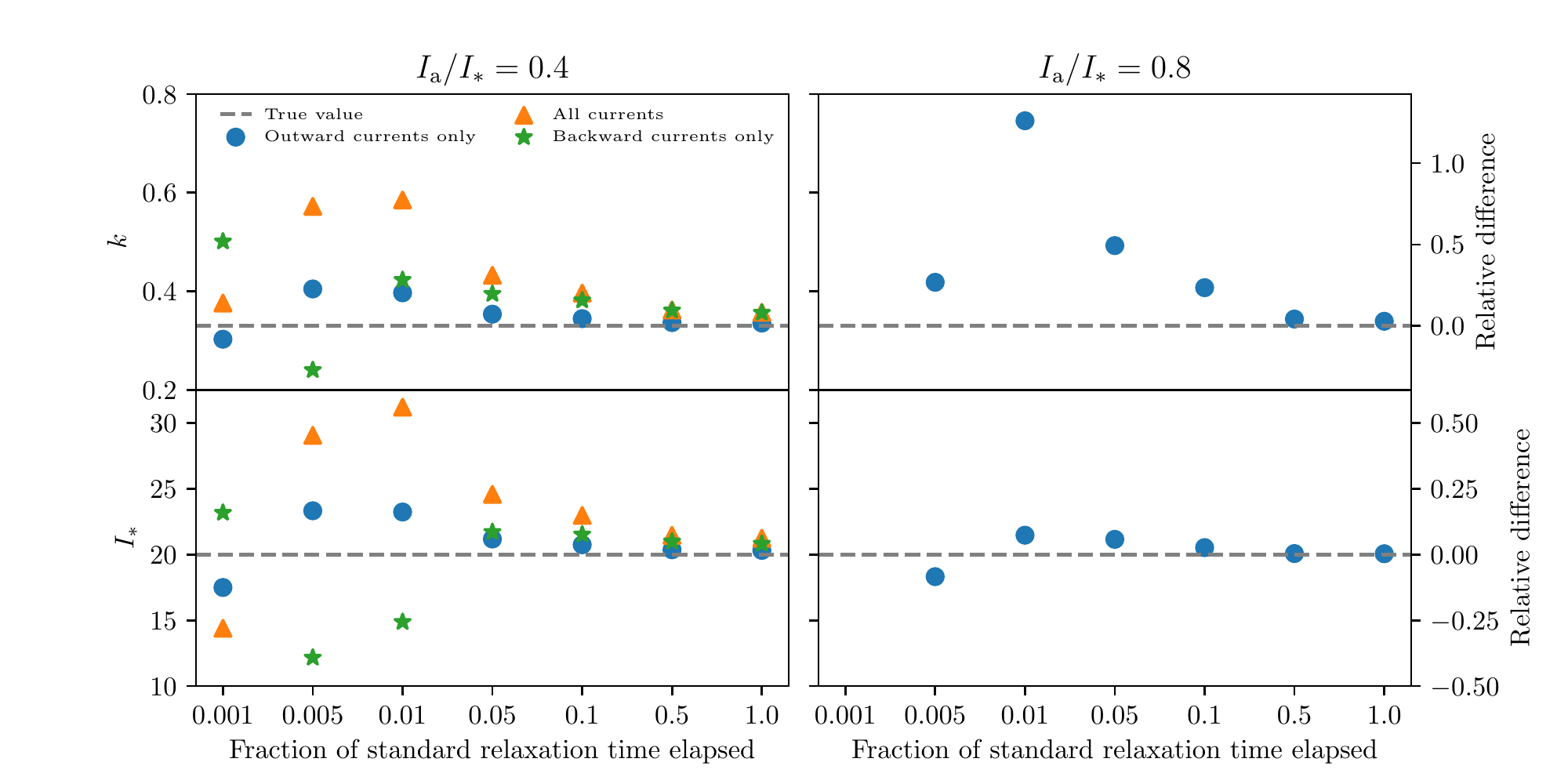}
    \caption{Fit results for $I_\ast$ and $\kappa$ as a function of the relaxation time $\Delta t$ for two values of $I_\mathrm{a}/I_\ast$, using different subsets of the numerical data. Left: for $I_\mathrm{a}/I_\ast=0.4$, a good reconstruction performance is observed even for short fractions of the ideal relaxation time. A rather similar performance is observed for the three types of analysis, the one based on the outward currents having the best performance. Right: for $I_\mathrm{a} / I_\ast=0.8$, only the cases corresponding to longer fractions of the relaxation time feature a good performance. Moreover, only the analysis based on the outward currents is displayed, as the other two feature either failures or large errors in the fit. (Simulation parameters: initial $I_\mathrm{a}/I_\ast=0.4$, boundary step $\Delta I =0.1 \sigma^2$, 10 repetitions of the three-step procedure, data up to a maximum current recovery of $90\%$).}
    \label{fig:all_different_time}
\end{figure*}

After each execution of the proposed three-step protocol described in Section~\ref{sec:moving_the_absorbing_barrier} and shown in Fig.~\ref{fig:9}, we end up with two outward and one inward recovery currents. In every configuration explored, we observe better reconstruction results when considering only the recovery currents from the outward step. On the other hand, considering only the inward recovery currents or all currents simultaneously, poorer performance and numerical instabilities are observed. This is explained by the fact that when the position of the absorbing boundary is moved inwards, we are cutting in a distribution that is not necessarily in the perfect equilibrium configuration defined in our approximations. On the other hand, when we move the boundary condition outwards, we obtain a much more reliable observable of the distribution that populates the new available action interval, when evolving towards the new equilibrium state. It is possible to observe this behaviour in Fig.~\ref{fig:all_different_time}, where the relative error in the reconstruction of $\kappa$ and $I_\ast$ is shown for the three types of fit as a function of the fraction of the ideal relaxation time $\Delta t$ for two values of $I_\mathrm{a}/I_\ast$, representing the inner part of the stable-core region (left) and close to the change of regime of $D(I)$ (right). In the inner region, even small fractions of the relaxation time provide a good reconstruction of the fit parameters. Furthermore, the three types of analysis, based on outward only, inward only, and inward and outward recovery currents provide results with comparable accuracy, at least for longer fractions of the relaxation time.  

In the transition region, however, only longer fractions of the relaxation time provide a good reconstruction of the fit parameters and the only applicable type of analysis is the one based on the outward recovery currents. The other two analyses feature either failures or larger errors in the fit. Based on the observed behaviour, in the next plots we will only display results from the reconstruction based on the outward recovery currents.

\begin{figure*}[htp]
    \centering
    \includegraphics{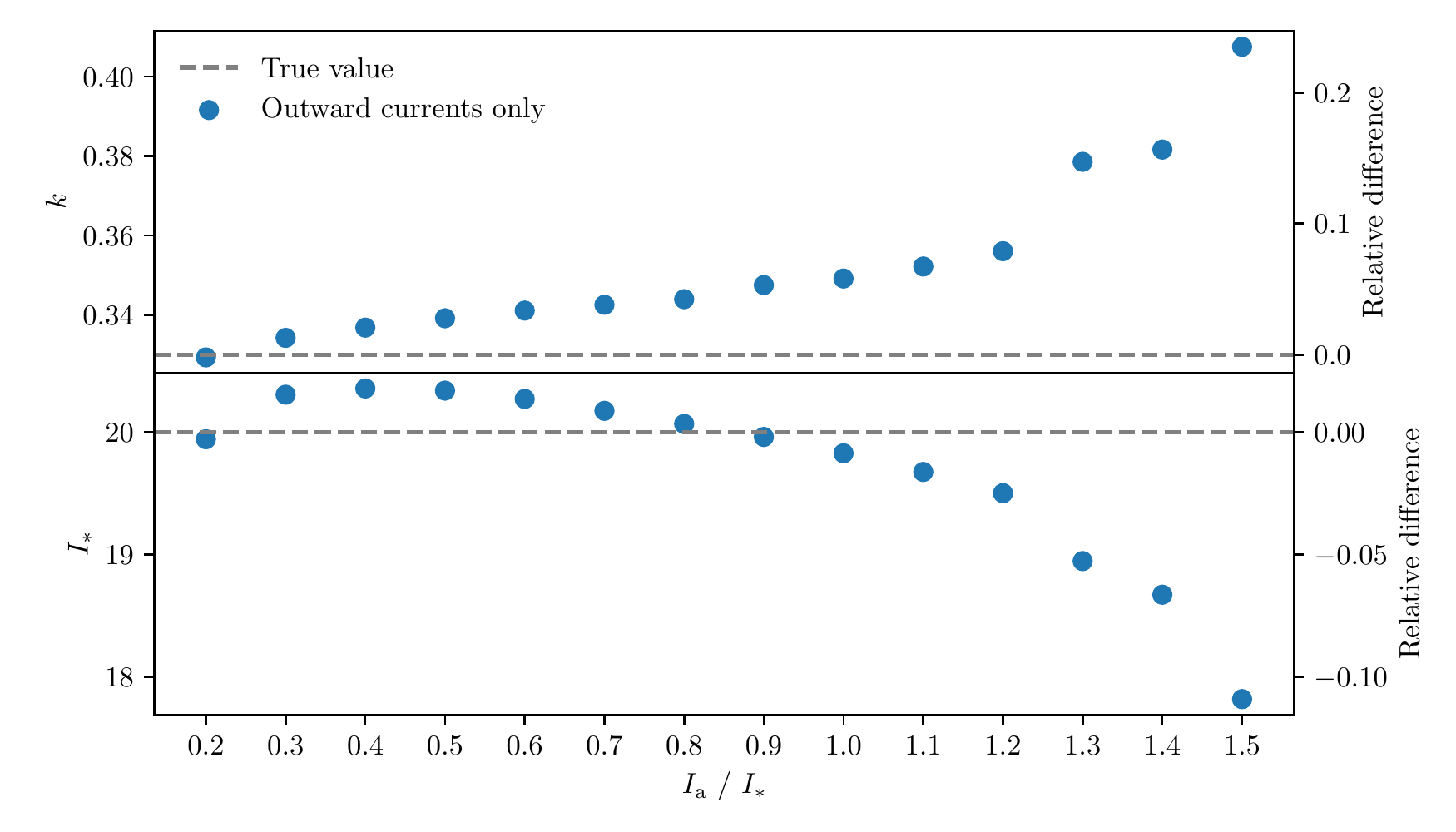}
    \caption{Fit results for $I_\ast$ and $\kappa$ as a function of the initial value of $I_\mathrm{a}/I_\ast$. The reconstruction performance is excellent for $I_\mathrm{a}/I_\ast \leq 1$, while for larger values the relative error increases. (Simulation parameters: boundary step $\Delta I=0.1 \sigma^2$, relaxation time $\Delta t=0.5$, 10 repetitions of the three-step procedure, data up to a maximum current recovery of $90\%$).}
    \label{fig:different_position}
\end{figure*}

In Fig~\ref{fig:different_position}, the reconstruction error for the two fit parameters as a function of the starting position of the absorbing boundary $I_\mathrm{a}/I_\ast$ is shown. 
The performance is excellent for $I_\mathrm{a}/I_\ast \leq 1$, whereas outside this region the reconstruction error increases. This behaviour is to be expected, as a recovery current carries mainly local information about $D(I)$. Hence, if only the region $I_\mathrm{a}/I_\ast > 1$ is sampled, the information about $D(I)$ reflects only its quasi-linear regime (see Fig.~\ref{fig:1}). Such incomplete information inevitably affects the performance of the final fit, as it prevents an accurate reconstruction of the strongly non-linear part of the diffusion coefficient. However, this result suggests also that, after having determined the fit parameters, one can verify whether the action interval explored was suitable for an accurate reconstruction of the functional form of the diffusion coefficient.

The plots in Fig.~\ref{fig:different_position} provide some insight on the dependence of the reconstruction performance as a function of a single parameter while the others are kept fixed. In the following figures, however, the relative error of the reconstruction procedure is shown as a function of two parameters. The colour code represents the relative difference between the reconstructed (from the proposed protocol) and the true (used in the numerical simulations of the FP processes) values of $I_\ast$ and $\kappa$ that describe $D(I)$. The scale is limited to a $20\%$ relative difference, which is assumed as a threshold to identify a poor performance. Note that white cells represent cases in which the reconstruction procedure failed.

\begin{figure*}[htp]
    \centering
    \includegraphics[width=\textwidth]{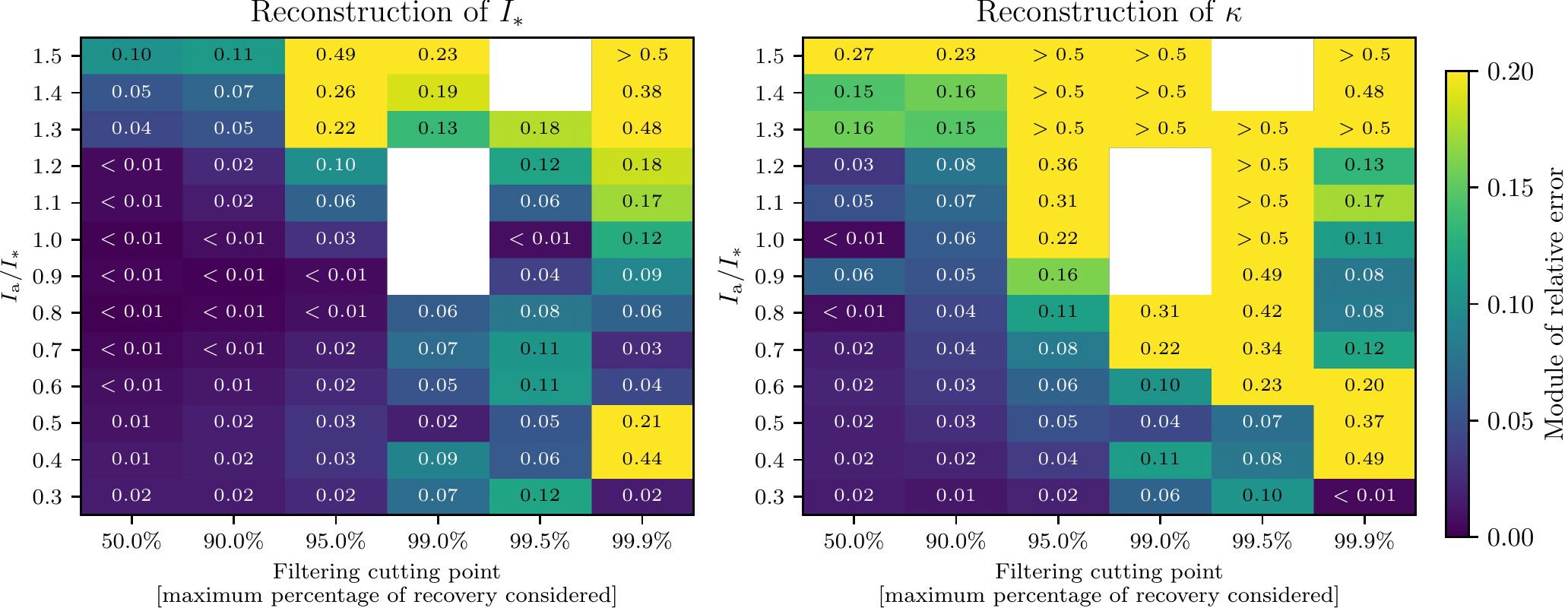}
    \caption{2D view of the reconstruction performance as a function of the cut in the recovery current, as applied in Fig.~\ref{fig:postprocessing}, and of the initial value of $I_\mathrm{a}/I_\ast$. It is clearly seen how certain values of the cut provide a consistent increase in reconstruction performance. White regions indicate a failure in convergence in the final fit procedure. (Simulation parameters: relaxation time $\Delta t=1$, boundary step $\Delta I=0.1 \sigma^2$, 10 repetitions of the three-step, data up to a maximum current recovery of $90\%$).}
    \label{fig:different_cut}
\end{figure*}

Figure~\ref{fig:different_cut} shows the relative error as a function of the cut of the recovery current performed during the post-processing and $I_\mathrm{a}/I_\ast$. The performance of the reconstruction approach improves when the recovery currents are cut. This depends on the fact that our approach is local, i.e.\ accurate to describe the system's behaviour close to the boundary condition. The longer the recovery time, the less local is the information gathered from the current. For this reason, the plots concerning the reconstruction performance are based on recovery currents cut at $90\%$. 

\begin{figure*}[htp]
    \centering
    \includegraphics[width=\textwidth]{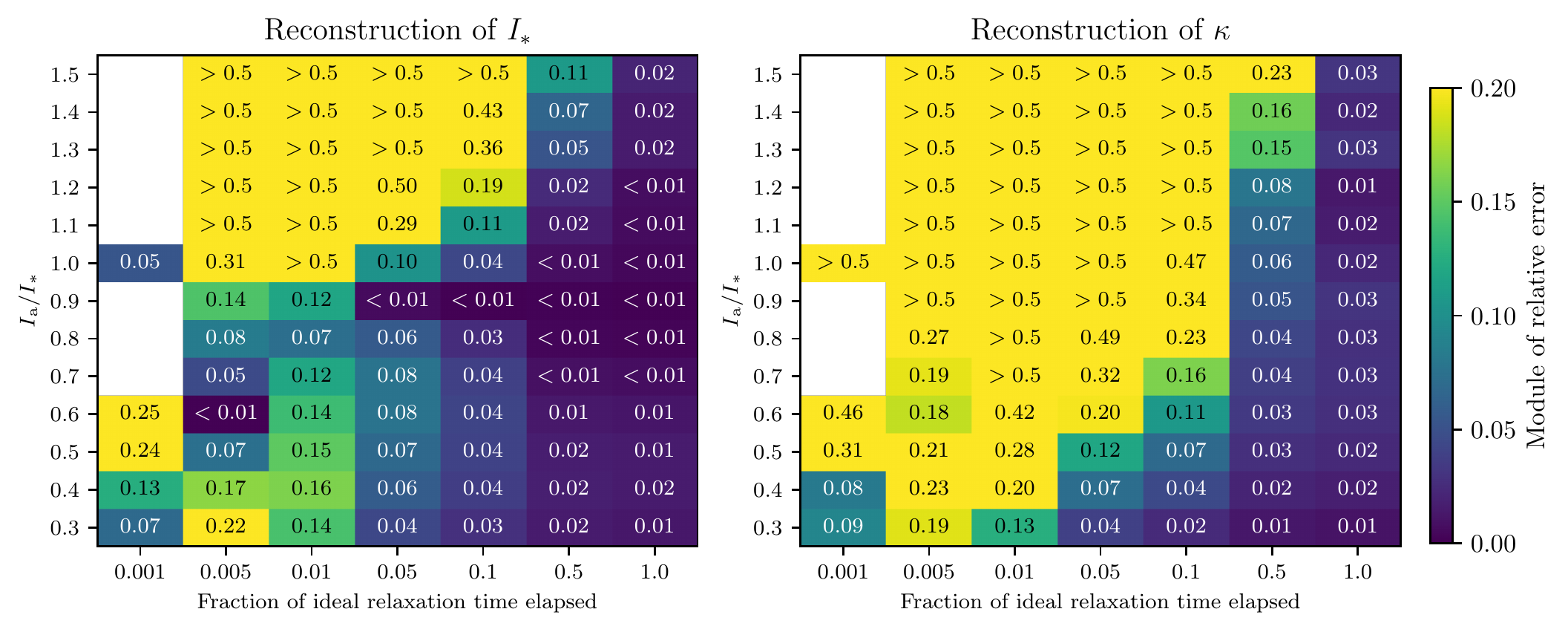}
    \caption{2D view of the reconstruction performance as a function of the relaxation time $\Delta t$ and the initial value of $I_\mathrm{a}/I_\ast$. It is clearly seen how the reconstruction performance improves for longer relaxation times. However, the method proves to be rather robust for fractions of ideal time up to $5\%$ provided $I_\mathrm{a}/I_\ast < 0.6$. White regions indicate a failure in convergence in the final fit procedure. (Simulation parameters: boundary step $\Delta I=0.1 \sigma^2$, 10 repetitions of the three-step procedure, data up to a maximum current recovery of $90\%$, if smaller values of $\Delta t$ lead to a normalised recovery below $90\%$, all the normalised recovery current data are used for the reconstruction).}
    \label{fig:different_time}
\end{figure*}

As for the reconstruction performance as a function of the  relaxation time $\Delta t$, Fig.~\ref{fig:different_time} shows the behaviour including the dependence on $I_\mathrm{a}/I_\ast$. In this case, the longer the relaxation time, the better is the performance of the reconstruction. In particular, longer relaxation times allow a better reconstruction even for large values of $I_\mathrm{a}/I_\ast$. It is worth highlighting that in the case of short relaxation times, a good overall performance can be achieved only by working at $I/I_\ast \ll 1$.
Combining the results of the last two analyses, one concludes that the best approach for an accurate determination of $I_\ast$ and $\kappa$ consists in increasing the relaxation time between successive changes of the position of the boundary condition and cutting the data from the recovery currents. 

\begin{figure*}[htp]
    \centering
    \includegraphics[width=\textwidth]{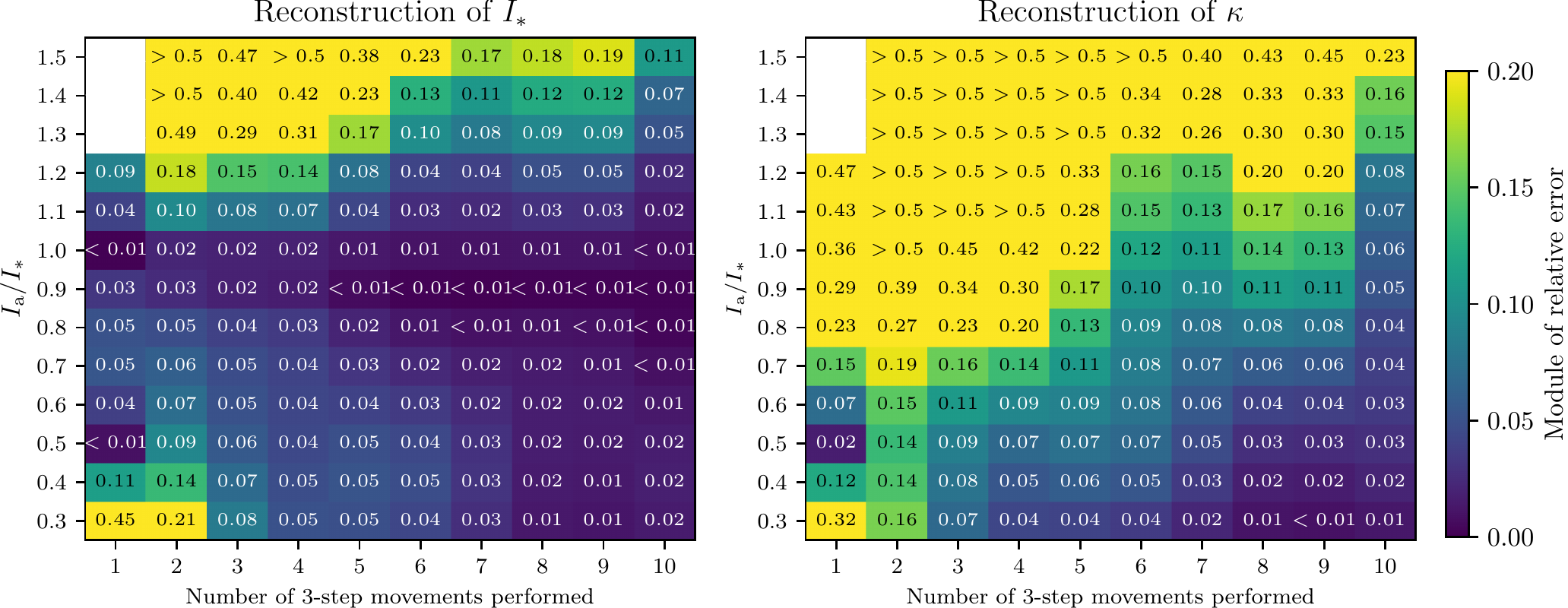}
    \caption{2D view of the reconstruction performance as a function of the number of three-step movements and of $I_\mathrm{a}/I_\ast$ starting positions. It is clearly seen how the performance increases with the number of three-step movements, as larger regions of the phase space are explored (as in Fig.~\ref{fig:9}). White regions indicate a failure in convergence in the final fit procedure. (Simulation parameters: relaxation time $\Delta t=0.5 \, [\text{a. u.}]$, boundary step $\Delta I=0.1 \sigma^2$, which is $\Delta I / I_\ast = 0.005$, data up to a maximum current recovery of $90\%$ is considered).}
    \label{fig:different_nsamples}
\end{figure*}

Figure~\ref{fig:different_nsamples} shows the 2D plot of the reconstruction performance as a function of the number of repetitions of the three-step procedure and $I_\mathrm{a}/I_\ast$. It can be seen with the higher number of repetitions, how the performance improves. This is naturally linked to the fact that repeating the three-step procedure implies sampling a larger extent of phase space, thus probing more accurately the behaviour of the diffusion coefficient as a function of the action. It is also clearly visible that starting from six repetitions of the three-step procedure, a good reconstruction is obtained for $I_\mathrm{a}/I_\ast < 1$.

\begin{figure*}[htp]
    \centering
    \includegraphics[width=\textwidth]{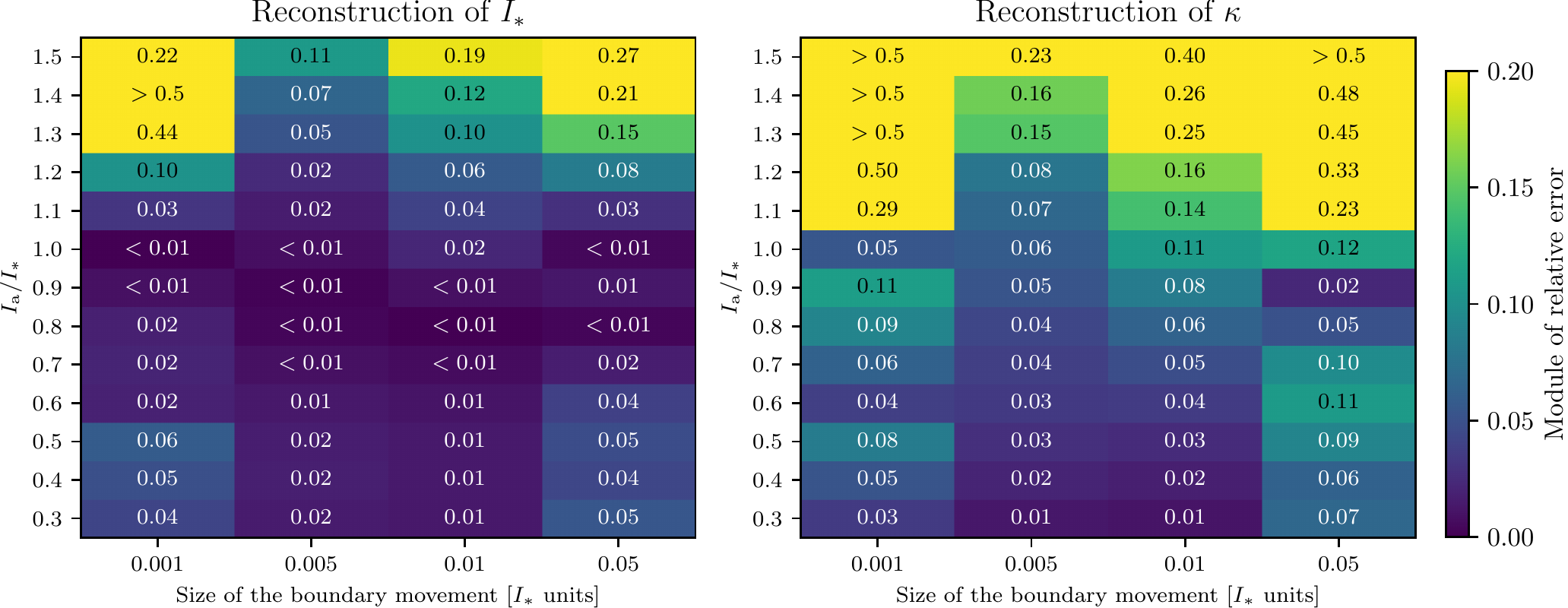}
    \caption{2D view of the reconstruction performance as a function of $\Delta I$ and $I_\mathrm{a}/I_\ast$. The  best performance is achieved for $\Delta I = 0.005 \sigma^2$, but the dependence on $\Delta I$ is very weak (Simulation parameters: relaxation time $\Delta t=0.5 \, [\text{a. u.}]$, 10 repetitions of the three-step procedure, data up to a maximum current recovery of $90\%$).}
    \label{fig:different_movement_module}
\end{figure*}

Finally, the impact of $\Delta I$ is shown in Fig.~\ref{fig:different_movement_module}, where the performance as a function of $\Delta I$ and $I_\mathrm{a}/I_\ast$ is depicted.  We see how the performance is not strongly affected by the choice of $\Delta I$, i.e.\ relative error fluctuations are less than $10\%$ for differences of an order of magnitude in $\Delta I$. However, it is important to highlight two facts that might suggest a choice in the size of the change of position of the absorbing boundary condition: (1) the ideal relaxation time is directly proportional to the size of the absorbing boundary movement; (2) a too small $\Delta I$ might lead to a too local sampling in action space, thus negatively affecting the final reconstruction of $D(I)$. 

%%%%%%%%%%%%%%%%%%%%%%%%%%%%%%%%%%%%%%%%%%%%%%%%%%%%%%%%%%%%%%%%%%%%%%%%%%%%%%%%

\section{Conclusions}
\label{sec:conclusions}

%%%%%%%%%%%%%%%%%%%%%%%%%%%%%%%%%%%%%%%%%%%%%%%%%%%%%%%%%%%%%%%%%%%%%%%%%%%%%%%%

Beam-halo scans, performed with movable collimators jaws, have been used intensively for probing the diffusive behaviour of the beam halo in circular accelerators and seems a very useful tool for probing this special regime of beam dynamics in the absence of beam instrumentation capable of providing diagnostic tools to study beam-halo dynamics.

In this paper, we have presented that, starting from a general framework to describe the dynamics of stochastic Hamiltonians, it is possible to derive a Fokker-Planck equation with a diffusion coefficient whose functional form is based on the time stability estimate from Nekhoroshev theorem. The main result is the identification of an efficient protocol for probing the shape of such a  diffusion coefficient. The protocol has been scrutinised by means of detailed numerical simulations, and it is clear that eventually it should be tested with beam measurement data. Two aspects should be highlighted: although the framework presented in this article is one-dimensional, i.e.\ considering non-linear beam dynamics in one degree of freedom, we believe that it can be applied to cases representing systems with two degrees of freedom (as shown in Ref.~\cite{bazzani2020diffusion}). Furthermore, the proposed approach does not rely on a previous knowledge of the beam distribution, which is a clear advantage for applications. 
 
The proposed protocol relies on the idea that it is possible to separate the measured outgoing current into a global current, i.e.\ the general outgoing current loss that is measured from the exponentially slow erosion of the stable core of the beam, and a recovery current, i.e.\ the current following a change of the position of the boundary condition, which corresponds to a non-equilibrium state. By performing an alternating three-step sequence of outward-inward-outward boundary-condition changes, which can easily be done by means of collimator scans, it is possible to reconstruct the global current of the erosion process and use that to normalise the recovery currents. Each normalised recovery current ultimately contains local information on the diffusion coefficient without the need of prior knowledge on the form of the initial distribution in action space, and can be used for estimating its global shape.

The performance of this protocol has been tested by means of a large number of simulated Fokker-Planck processes performed in various configurations, to evaluate the reliability and limits of our approach. The protocol proved to be capable of reconstructing with precision and good accuracy the parameters of the diffusion coefficient when it is performed in a phase-space region where the diffusion coefficient has an exponential evolution, i.e.\ for $I/I_\ast < 1$, the relaxation time between boundary condition changes is long enough so that the system reaches an equilibrium state, and multiple amplitudes have been probed. For this last condition, the optimal number of amplitudes to be sampled is highly dependent on the detail of the diffusion process, however, from the simulations it appears that about six sequences of three-step absorbing boundary changes covering the $I / I_\ast < 1$ region is a good choice. 

The analysis also highlighted how a good reconstruction performance can be achieved by considering only the outgoing recovery currents in the final fitting reconstruction, and by discarding part of the recovery current data beyond a certain level, as it is more prone to reconstruction errors and more difficult to characterise with our analytical formulas. It is worth stressing that the reconstruction performance proves to be good even if the optimal conditions are not fully met. Most importantly, the procedure provides useful information about possible shortcomings present in the dataset under consideration, such as a high uncertainty band in the global current reconstruction, or a reconstructed value of $I_\ast$ that indicates that the probed phase-space region is outside of the optimal interval $I / I_\ast < 1$. In these cases, the protocol should be reapplied in better conditions, e.g.\ by adjusting the range of actions probed to satisfy the condition $I / I_\ast < 1$.

Thanks to the positive and encouraging results of the analysis presented here, we are confident that the measurement protocol is a powerful tool for probing the non-linear diffusive behaviour in an accelerator like the LHC. As a future step, the protocol will be applied to the available beam-halo collimator-scan data collected at the LHC, to attempt an improved reconstruction of the diffusion coefficient with respect to previous analyses. Further to this, a proposal of a dedicated measurement at the LHC, performed in the optimal conditions considered in this paper, will complete these investigations. 

%%%%%%%%%%%%%%%%%%%%%%%%%%%%%%%%%%%%%%%%%%%%%%%%%%%%%%%%%%%%%%%%%%%%%%%%%%%%%%%%
%\section*{Data availability}
%
%Data sharing not applicable to this article as no datasets were generated or analysed during the current study.
%

% \clearpage

\appendix

%%%%%%%%%%%%%%%%%%%%%%%%%%%%%%%%%%%%%%%%%%%%%%%%%%%%%%%%%%%%%%%%%%%%%%%%%%%%%%%%
 
\section{Numerical integration of the Fokker-Planck equation using the Crank-Nicolson method} \label{app_sec:numerical_integration_with_crank_nicolson}

%%%%%%%%%%%%%%%%%%%%%%%%%%%%%%%%%%%%%%%%%%%%%%%%%%%%%%%%%%%%%%%%%%%%%%%%%%%%%%%%

For executing the numerical integration of a FP equation in the form of Eq.~\eqref{eq:fp}, we used the Crank-Nicolson integration scheme~\cite{crank1947practical}, which is a finite difference method, second-order and implicit in time. It can be shown that this scheme is unconditionally stable for many differential equations~\cite{thomas2013numerical}.

To obtain valid numerical results of the integration of the FP equation with Nekhoroshev-like diffusion coefficient like in Eq.~\eqref{eq:diffusion}, and obtain a consistent evaluation of the outgoing current in different scenarios, we have to properly evaluate the stiffness of the problem and adapt consequently the fineness in both time and space discretisations. Moreover, concerning the simulation of an instantaneous change in position of the absorbing boundary condition, a rigorous protocol must be established, especially when considering the inward displacements of the boundary condition, for which some additional precautions must be taken.

The outgoing current, defined in Eq.~\eqref{eq:outgoing_current_definition}, is obtained by computing directly, in between each integration step, the numerical derivative of $\rho$ at the absorbing boundary position. 

A Nekhoroshev-like diffusion coefficient has the main characteristic of varying by various orders of magnitude over the accessible range of the the action variable, meaning that if we want to simulate the entirety of a diffusive phenomenon, we must take into consideration such a wide range of values in the integration process. This becomes mostly critical when the process to be simulated is the recovery current that occurs after a variation of the position of the boundary condition, like the ones described in Section~\ref{sec:moving_the_absorbing_barrier}.

The recovery current is mainly dependent on variations in the equilibrium distribution that are various orders of magnitude lower, in absolute value, than the core part of the distribution (refer to Fig.~\ref{fig:3} and~\ref{fig:5}). It is therefore necessary to choose a time and space discretisation fine enough to obtain numerical estimates that are not seriously affected by the integration error. To do that, we performed a convergence test for a single recovery current in every scenario we wanted to analyse. In such a convergence test, we increased gradually the fineness of the discretisations, until we measured a relative difference between numerical results not higher than $1\%$.

When it comes instead to reproduce the instantaneous change of the position of the absorbing boundary condition in the integration scheme, we perform a re-sampling of the distribution $\rho$ at the time of the boundary change, while keeping the same fineness for the spatial discretisation. In an outward movement, however, this process is straightforward, as there is no artificial change to the existing distribution $\rho$ to be taken into account, and we just add an empty region with no singular points. Instead, for the case of an inward movement, we do have to perform a cut inside the $\rho$ distribution, corresponding to the movement performed by the absorbing boundary. Such a cut generates an inconsistency between the non-zero value of $\rho$ at the new position of the boundary condition and the zero condition imposed by the absorbing boundary condition. This inconsistency leads to a divergence in the analytical definition of the outgoing current and undefined behaviours in the numerical integration. Therefore, we apply to the cut distribution a sharp damping, right next to the newly positioned absorbing boundary condition, generated by a logistic function $f(I)$ defined as
\begin{equation}
    f(I) = \frac{1}{1 + e^{\frac{I-I_\text{a}+\ell}{\ell}}} \, , 
    \label{eq:logistic_damping}   
\end{equation}
where $\ell$ is the extent of the range of action values where the damping occurs, and is taken equal to two twice the fineness of the spatial discretisation, and $I_\text{a}$ is the position of the absorbing boundary condition after the inward movement. In this way, $\rho_d(I) = \rho(I) f(I)$ represents a distribution that is smooth enough to avoid instabilities in the numerical integration. The sharpness of this damping is directly proportional to the fineness of the spatial sampling, and its effects are included in the convergence tests.

%%%%%%%%%%%%%%%%%%%%%%%%%%%%%%%%%%%%%%%%%%%%%%%%%%%%%%%%%%%%%%%%%%%%%%%%%%%%%%%%

\section{Analytical estimate of the outgoing current for a FP process}
\label{app_sec:analytic_estimate_of_the_current_loss}

%%%%%%%%%%%%%%%%%%%%%%%%%%%%%%%%%%%%%%%%%%%%%%%%%%%%%%%%%%%%%%%%%%%%%%%%%%%%%%%%

We are interested in finding a good analytical approximation for the outgoing current of a FP process like Eq.~\eqref{eq:fp}. We start  by applying the following change of variables
\begin{equation}
    x = -\int_I^{I_\mathrm{a}} \frac{1}{D^{1/2}(I')}\,\mathrm{d}I'\,,\quad \rho_x(x,t)=\rho(I,t)\frac{\mathrm{d}I}{\mathrm{d}x}=\rho(I,t)\sqrt{D(I)} \, ,
\end{equation}
which leads to
\begin{equation}
    \pdv{\rho_x}{t} = \frac{1}{2}\pdv{x} \left[\frac{1}{D^{1/2}}\dv{D^{1/2}}{x}\rho_x\right]+\frac{1}{2}\pdv[2]{\rho_x}{x}\, ,
\end{equation}
where $D=D\left(I(x)\right)$. By introducing the effective potential $V(x)=-\ln(D^{1/2}(x))$, we obtain the Smoluchowsky form~\cite{hannes1996fokker}
\begin{equation}
    \pdv{\rho_x}{t} = \frac{1}{2}\pdv{x} \dv{V(x)}{x} \rho_x+\frac{1}{2}\pdv[2]{\rho_x}{x}\, .
    \label{eq:smol}
\end{equation}
Equation~\eqref{eq:smol} can be made self-adjoint by means of the following change of variables 
\begin{equation}
    %\rho'(x,\tau) = \exp\left[-\frac{V(x)}{2D}\right]p(x,\tau) \, ,
    \rho_x(x,t) = \exp\left[-\frac{V(x)}{2}\right]p(x,t) \, ,
\end{equation}
and Eq.~\eqref{eq:smol} is cast into the following form
\begin{equation}
    \pdv{p}{t} = \frac{1}{4}\left[\dv[2]{V}{x} - \frac{1}{2}\left(\dv{V}{x}\right)^2\right]p + \frac{1}{2}\pdv[2]{p}{x}\,.
    \label{eq:self-adj}
\end{equation}

The general solution of Eq.~\eqref{eq:self-adj} can be written as
\begin{equation}
    p(x,t) = \sum_\lambda c_\lambda(t)\phi_\lambda(x)\,,
    \label{eq:expansion}
\end{equation}
where an expansion using the eigenfunctions $\phi_\lambda(x)$ of the operator on the r.h.s.\ of Eq.~\eqref{eq:self-adj} has been used, namely 
\begin{equation}
    2\left\{-\frac{1}{4}\left[\dv[2]{V}{x} - \frac{1}{2}\left(\dv{V}{x}\right)^2\right] - \lambda \right\}\phi_\lambda(x) = \dv[2]{\phi_\lambda}{x}\,,
    \label{eq:eigenproblem}
\end{equation}
and $c_\lambda(t) = c_\lambda(0)e^{-\lambda t}$. This choice of eigenfunctions is motivated by the working hypothesis that $p(x,t\to+\infty)=0$, i.e.\ the system will eventually relax to a zero distribution.

By using the orthogonality and completeness properties of $\phi_\lambda(x)$,
\begin{align}
    \int \phi_\lambda(x)\phi_\lambda(x')\,\mathrm{d}x &= \delta(\lambda - \lambda')\\
    \sum_\lambda \phi_\lambda(x)\phi_\lambda(x') &= \delta(x-x')\,,
\end{align}
and considering the initial condition
\begin{align}
    \rho_x(x,0) &= \exp\left[-\frac{V(x)}{2}\right] p(x,0) \\
    p(x,0) &= \sum_\lambda c_\lambda(0) \phi_\lambda(x)\,, \label{eq:mid_step_analytic}
\end{align}
we have that
\begin{align}
    c_\lambda(0) = \int \exp\left[\frac{V(x)}{2}\right] \rho_x(x,0)\phi_\lambda(x)\, \mathrm{d}x\,,
    \label{eq:c_lambda_zero}
\end{align}
and the solution for an initial Dirac delta distribution $\rho_x(x, 0)=\delta(x-x_0)$ can be written as
\begin{equation}
    \rho_x(x,t)=\exp\left[\frac{V(x_0)-V(x)}{2}\right]\sum_\lambda e^{-\lambda t}\phi_\lambda(x_0)\phi_\lambda(x)\, ,
\end{equation}
and the outgoing current at an absorbing boundary in $x=0$, which in the original variables corresponds to $I=I_\mathrm{a}$, reads
\begin{equation}
    J(t) = \frac{1}{2}\pdv{\rho_x}{x}\Bigr|_{(0,t)}     \, .
    \label{eq:eigencurrent}
\end{equation}

If the potential is linearised, i.e.\ $V(x)\simeq -\nu \, x$, then there is an analytic solution to the eigenvalue problem in Eq.~\eqref{eq:eigenproblem}
\begin{equation}
    -2\left[\lambda-\frac{\nu^2}{2}\right]\phi_\lambda(x)=\dv[2]{\phi_\lambda}{x}\, ,
\end{equation}
and if we replace this solution in Eq.~\eqref{eq:eigencurrent}, we obtain the expression for the outgoing current
\begin{equation}
    J(x_0, t) = \frac{|x_0|}{t\sqrt{2\pi t}}\exp\left(-\frac{(x_0+\frac{\nu}{2}t)^2}{2t}\right) \,,
    \label{eq:out_current}
\end{equation}
which has dimension $t^{-1}$. Furthermore, the linearisation $\nu$ of the potential $V(x)$ near $x=x_0$ reads
\begin{equation}
    \nu=\frac{\frac{1}{2\kappa}}{I(x_0)}\left(\frac{I_\ast}{I(x_0)}\right)^{\frac{1}{2\kappa}}\exp\left[-\left(\frac{I_\ast}{I(x_0)}\right)^{\frac{1}{2\kappa}}\right]\,,
\end{equation}
which can be inserted into Eq.~\eqref{eq:thecurrent}, for obtaining an analytical estimate of the outgoing current.

%%%%%%%%%%%%%%%%%%%%%%%%%%%%%%%%%%%%%%%%%%%%%%%%%%%%%%%%%%%%%%%%%%%%%%%%%%%%%%%%

\section{Outgoing current for a system with infinite source}\label{app_sec:outgoing_current_for_a_system_with_infinite_source}

%%%%%%%%%%%%%%%%%%%%%%%%%%%%%%%%%%%%%%%%%%%%%%%%%%%%%%%%%%%%%%%%%%%%%%%%%%%%%%%%

To make use of the analytical  estimate of the outgoing current presented in Appendix~\ref{app_sec:analytic_estimate_of_the_current_loss}, we need to slightly modify certain steps to adapt to the different non-zero equilibrium distribution $\rho_\text{eq}$, as the original calculations are carried out under the assumption that $\rho(I,t\to+\infty)=0$, and modifications to Eq.~\eqref{eq:mid_step_analytic} need to be made and then propagated.

Under these new conditions, the expansion of the solution of the diffusive problem in Eq.~\eqref{eq:expansion} can be modified according to 
\begin{equation}
    p(x, t) = \sum_\lambda c_\lambda(t) \phi_\lambda(x) + \exp\left[\frac{V(x)}{2}\right] \rho'_\text{eq}(x)\,,
    \label{eq:new_expansion}
\end{equation}
where $\rho'_\text{eq}(x) = \rho_\text{eq}(I(x))\dv{I}{x}$ is the equilibrium distribution of our system, while considering the change of variables necessary to work with the self-adjoint diffusive problem in the Smoluchowsky form. The various considerations about $c_\lambda(t)$ and $\phi_\lambda(x)$ are unchanged. The values $c_\lambda(0)$ should be recomputed and from the expansion in Eq.~\eqref{eq:new_expansion}, we obtain
\begin{align}
    \rho'(x, 0) &= \exp\left[-\frac{V(x)}{2}\right]p(x,0)\\
    p(x, 0) &= \sum_\lambda c_\lambda(t) \phi_\lambda(x) + \exp\left[\frac{V(x)}{2}\right] \rho'_\text{eq}(x)\,,
\end{align}
which then leads to
\begin{equation}
    c_\lambda(0) = \int \exp\left[\frac{V(x)}{2}\right] \left\{\rho'(x,0) - \rho'_\text{eq}(x)\right\}\phi_\lambda(x)\, \mathrm{d}x = \int \exp\left[\frac{V(x)}{2}\right] \rho^\ast(x,0)\phi_\lambda(x)\, \mathrm{d}x\,,
\end{equation}
where, $\rho^\ast(x,t)$ stands for the difference between the actual and the equilibrium distribution, still to be reached, and in this framework, the rest of the analytic current estimate, i.e.\ Eq.~\eqref{eq:out_current}, still applies.

%%%%%%%%%%%%%%%%%%%%%%%%%%%%%%%%%%%%%%%%%%%%%%%%%%%%%%%%%%%%%%%%%%%%%%%%%%%%%%%%
%
\clearpage
\bibliographystyle{unsrt}
\bibliography{CEMbibliography.bib}
%
%%%%%%%%%%%%%%%%%%%%%%%%%%%%%%%%%%%%%%%%%%%%%%%%%%%%%%%%%%%%%%%%%%%%%%%%%%%%%%%%

\end{document}